%% file: main.tex
\documentclass{article}
\PassOptionsToPackage{table,xcdraw}{xcolor}   
\usepackage{arxiv}
\usepackage[utf8]{inputenc}
\usepackage[T1]{fontenc}
\usepackage{hyperref}
\usepackage{url}
\usepackage{booktabs}
\usepackage{amsfonts}
\usepackage{microtype}
\usepackage{graphicx}
\usepackage{caption}
\graphicspath{ {./figures/final/} }
\usepackage{makecell}
\usepackage[numbers]{natbib}

\usepackage{amsmath,amssymb}
\usepackage{subcaption}
\usepackage{enumitem}
\usepackage{algorithmic}
\usepackage{textcomp}
\usepackage{tabularx}
\usepackage{ragged2e}
\usepackage{adjustbox}
\usepackage{multirow}
\usepackage{rotating}
\usepackage{array}
\usepackage{comment}
\usepackage{tcolorbox}
\usepackage[english]{babel}
\usepackage{tikz}
\usepackage[table,xcdraw]{xcolor}   
\usepackage{lscape}
\usepackage{orcidlink}              
\definecolor{springernatureblue}{HTML}{004BA0} 
\definecolor{figure-navy-blue}{HTML}{004B78}
\definecolor{figure-soft-white}{HTML}{F5F5F5}

\title{Fine-Tuning Large Language Models to Classify Pull Request--Issue Alignments: Going Beyond Prompting}

\renewcommand{\headeright}{}
\renewcommand{\undertitle}{}
\author{
  Mustafa Yasir Altunhan\thanks{These authors contributed equally.} \\{}
  Department of Computer Engineering \\{}
  Bilkent University \\{}
  Ankara, Turkey \\
  \texttt{yasir.altunhan@bilkent.edu.tr} \\
  \href{https://orcid.org/0000-0002-4584-5643}{ORCID: 0000-0002-4584-5643}
  \And
  H\"{u}seyin \"{O}zg\"{u}r Kamal{\i}\footnotemark[1] \\{}
  Department of Computer Engineering \\{}
  Bilkent University \\{}
  Ankara, Turkey \\
  \texttt{ozgurkamali@bilkent.edu.tr} \\
  \href{https://orcid.org/0009-0009-9864-9513}{ORCID: 0009-0009-9864-9513}
  \And
  Eray T\"{u}z\"{u}n \\{}
  Department of Computer Engineering \\{}
  Bilkent University \\{}
  Ankara, Turkey \\
  \texttt{eraytuzun@cs.bilkent.edu.tr} \\
  \href{https://orcid.org/0000-0002-5550-7816}{ORCID: 0000-0002-5550-7816}
}

\begin{document}
\maketitle

\begin{abstract}
   
\noindent\textbf{Context} Accurate alignment between pull requests (PRs) and corresponding issues is crucial for efficient software development and maintaining code quality, as these misalignments can lead to reduced traceability, hindered defect localization, and decreased maintainability. 

\noindent\textbf{Objective} This study aims to improve automated PR–issue alignment classification by leveraging fine-tuned large language models (LLMs) across multiple alignment categories, and conducts interpretability analysis to investigate the effects of PR-issue fields on the fine-tuned LLMs predictions.

\noindent\textbf{Method} Our methodology consists of dataset preparation, LLM fine-tuning, and interpretability analysis. We first extended an existing dataset and applied data augmentation to address class imbalance. Subsequently, GPT-4o was fine-tuned via instruction-tuning, and open-source LLMs: including CodeLlama-7B, CodeQwen1.5-7B, StableCode-3B, CodeGemma-7B, and Deepseek-Coder-6.7B were fine-tuned using classification-specific heads. Additionally, interpretability analysis using Shapley Additive Explanations (SHAP) was conducted to examine the  influences of PR–issue fields on LLMs predictions for the best performing open-source LLM.

\noindent\textbf{Results} Fine-tuned LLMs outperformed baseline models, achieving average improvements of 6.15\% in accuracy and F1-micro, 14.69\% in F1-macro, and 6.15\% in recall. CodeLlama-7B emerged as the best-performing fine-tuned LLM overall, demonstrating consistent proficiency across metrics, while interpretability analysis revealed that code diffs together with issue body and PR body contents exert the greatest influence on predictions.

\noindent\textbf{Conclusions} Fine-tuning substantially enhances PR–issue alignment classification, improving both accuracy and efficiency. Furthermore, interpretability analysis provides actionable insights into the dataset features driving alignment decisions, deepening understanding of how LLMs reason over software artifacts.

\end{abstract}

\keywords{Pull Request-Issue Alignment \and Pull Request Intention \and Code Review \and Pull Request Review \and Large Language Models \and Fine-Tuning \and Instruction Tuning \and Interpretability \and SHAP \and Software Maintenance}
\input{introduction}

\input{related-work}

\input{methodology}
\input{results}
\input{threats-to-validity}
\input{discussion}

\input{conclusion}

\bibliographystyle{unsrtnat}
\bibliography{references}
\end{document}

%% file: introduction.tex
\section{Introduction}
\label{sec:introduction}
Pull Request (PR)-based development is a cornerstone of contemporary software engineering; however,  misalignment between PR commits and related issues can silently erode code traceability and maintainability \citep{herzig-msr-2013}. In modern software development environment developers open an issue that specifies a particular artifact or requirement, then a PR  is used to close the issue by addressing the specified artifact in the issue. This process is intended to promote structured collaboration, improve traceability and maintain consistency between code changes and documented requirements. Best practices suggest that each issue represents a single, well-defined task, and each PR should fully address the specific content of its corresponding issue \citep{microsoft2024pullrequests}. However, in practice, PR commits may contain changes that are not requested or discussed in the corresponding issues, a phenomenon known as tangled commits \citep{herbold2022fine}. Traditionally, these Tangling commits has been identified manually by developers, who take action if they think there is a irrelevant change. When Tangling commits are not adequately addressed, they can lead to reduced traceability, hindered defect localization, and decreased maintainability \citep{herbold2022fine}. Therefore, the effective detection and handling of Tangling is critical for ensuring project traceability and maintainability.

The problem of tangling is not merely a theoretical concern; its existence and detrimental effects have been widely validated by empirical studies. While various types inconsistencies exists between commits and issue, prior researchers have largely focused on tangling commits, a particularly pervasive and well-documented issue due to its measurable impact on software quality \citep{herzig-msr-2013,herbold2022fine}. A Tangling commit is defined as a commit that combines unrelated changes into a single commit. Tangling commits complicates code review and debugging processes by hindering program comprehension and eroding the separation of concerns. Studies have provided concrete evidence of this problem's prevalence and impact. For example, a seminal study by \citet{herzig-msr-2013} demonstrated that up to 20\% of bug-fixing commits are tangled, introducing substantial noise that can skew defect prediction models. Their findings revealed that this noise can lead to a considerable portion of files (up to 16.6\%) being incorrectly associated with bug reports. Building on this work, \citet{herbold2022fine} showed that depending on the context, up to 47\% of commits in open-source repositories involve task mixing, further highlighting the widespread nature of the problem. Such empirical findings validate the importance of the misalignment problem and demonstrate its widespread prevalence.

Previous studies have investigated methods to detect and untangle tangling commits. Some researchers focused on post-hoc detection and correction, while others explored preventive approaches. Building on the empirical evidence of tangling commits, researchers have developed automated and semi-automated methods to detect and untangle them. Early approaches, such as EpiceaUntangler \citep{Untangling-fine-grained-code-changes}, leveraged fine-grained edit histories from Integrated Development Environments (IDEs) to accurately cluster changes, though this method was limited by its heavy reliance on developer instrumentation. Subsequent efforts, including Flexeme \citep{flexeme} and ComUnt \citep{untangling-composite-commits-by-attributed-graph-clustering}, advanced the field by introducing more generalizable, graph-based techniques that framed the untangling problem as attributed graph clustering. While these approaches improved automation and applicability, they were often limited by the inherent challenges of unsupervised methods or their own performance. More recently, UTango \citep{utango} explored a supervised learning approach, demonstrating improved accuracy in detecting tangling commits by adapting to developer-specific grouping behaviors, but this came with the challenge of a high-quality labeled data requirement. In contrast to these post-hoc detection and untangling systems, other research has explored preventive approaches. For instance, \citet{are-you-committing-tangled-changes} proposed a method that analyzes past commits to learn common patterns and provides developers with real-time feedback to encourage more atomic commit practices.

Building on the limitations of prior work that primarily focused on the binary classification of tangled versus non-tangled commits, recent research by \citet{bilsen} extended this alignment to the relationship between PRs and issues, introducing a more nuanced taxonomy to capture the full spectrum of PR–issue alignment. Previous studies primarily focused on the existence of irrelevant concerns, but they did not systematically consider whether the intended issue was fully addressed \citep{Untangling-fine-grained-code-changes,flexeme,untangling-composite-commits-by-attributed-graph-clustering, utango}. This new taxonomy represents a significant step forward by formally defining and categorizing PR-issue alignment based on both the presence of irrelevant changes and the fulfillment of the issue's requirements. This work defines four distinct types: Exact, where a PR fully addresses an issue's requirements and includes no unrelated changes; Tangling, where a PR includes changes unrelated to the issue; Missing, where the PR does not fully address the intended issue; and Missing and Tangling, a combination of Missing and Tangling alignment. Furthermore, \citet{bilsen} used prompt engineering to assess LLMs performance in classifying PR–issue alignments according to this taxonomy.

Parallel to these research efforts, recent advancements in LLMs have significantly impacted text classification methodologies. LLMs like GPT-4 and Llama \citep{brown2020language,touvron2023llama} have enabled zero-shot and few-shot prompting, facilitating classification without extensive fine-tuning. However, while these models offer broad applicability, studies have consistently shown that their performance diminishes in domain-specific, fine-grained classification, particularly when processing noisy or highly technical software artifacts \citep{Gu_2025}. This limitation is especially problematic for tasks such as PR-issue alignment, where deep contextual understanding and high precision are essential.

Despite the advances in both detecting and untangling commits and the introduction of a new PR-issue alignment taxonomy \citep{bilsen}, A notable research gap remains in an effective classification method for PR–issue alignment. Previous work on tangling commits is limited, as it addresses the existence of irrelevant changes in commits and thus fails to measure whether the issue's requirements were actually satisfied \citep{Untangling-fine-grained-code-changes, flexeme,untangling-composite-commits-by-attributed-graph-clustering, utango}. Conversely, while the new taxonomy proposed by \citet{bilsen} provides a comprehensive framework, the prompt-based method used to evaluate it does not fully leverage the fine-tuning capabilities of LLMs and lacks the robustness and domain-specific precision required for this task. These gaps highlight the need for a new methodology that harnesses the full potential of LLMs to address the nuanced challenges of PR-issue alignment.

Our research aims to address these gaps by fine-tuning LLMs on a PR-issue dataset labeled with the new taxonomy classes. We compare the performance of these fine-tuned models with previous work to evaluate their effectiveness for this task \citep{bilsen}. Additionally, we analyze different fine-tuned LLMs to identify the most effective LLM in PR-issue alignment classification. Furthermore, our analysis investigates which parts of the PR–issue pair are most influential in models classification outcomes, providing insight into the LLMs decision-making process. To guide our research, we pose the following research questions:

\begin{enumerate}[label=\textbf{RQ\arabic*:}, leftmargin=*, align=left]
    \item \textbf{How do fine-tuned LLMs perform in classifying PR–issue alignment compared to previous approaches?} \\
    To answer this question, we apply supervised fine-tuning to both closed-source and open-source models, though with different training strategies. For GPT-4o, a state-of-the-art closed-source LLM, we leverage its generative capabilities by instruction tuning it to produce the correct class label through text generation. For the open-source models -CodeLlama-7B, CodeQwen1.5-7B, StableCode-3B, CodeGemma-7B, and Deepseek-Coder-6.7B- we fine-tune classification-specific heads that output probability distributions over the defined alignment categories. We then compare the classification performance and storage space requirements of these fine-tuned models against the prior prompt-based approach proposed by \citet{bilsen}.
    
    \item \textbf{Which fine-tuned LLM achieves the best performance in classifying PR–issue alignment?} \\
    To address this RQ, we compare the performance metrics of fine-tuned instances of GPT-4o, CodeLlama-7B, CodeQwen1.5-7B, StableCode-3B, CodeGemma-7B, and Deepseek-Coder-6.7B, rank each LLM on individual metrics, and compute an overall rank to identify the most effective fine-tuned LLM for PR–issue alignment classification.
    
    \item \textbf{Which PR-issue fields exert the greatest influence on classification prediction?} \\
    To answer this question, we apply Shapley Additive Explanations (SHAP) \citep{lundberg2017unified} analysis to the open-source LLM that achieved the highest accuracy to reveal the contribution of different fields within the PR-issue pairs to the LLMs classification decisions. 

\end{enumerate}

The rest of the paper is structured as follows: Section \ref{sec:related-work} reviews related literature. Section \ref{sec:methodology} describes the proposed methodology. Section \ref{sec:results} presents and interprets the experimental findings. Section \ref{sec:threats-to-validity} details the study's threats to validity and limitations. Section \ref{sec:discussion} discusses the results and suggests future research directions. Section \ref{sec:conclusion} concludes the paper.

%% file: related-work.tex
\section{Related Work}
\label{sec:related-work}

The existing body of research related to this study can be broadly investigated through five categories of work. The Section \ref{subsec:pr-issue-alignment}, provides an overview of empirical studies that have established the prevalence and impact of tangling commits and PR-issue misalignment. This is followed by a more focused discussion in Section \ref{subsec:tangling-commits}, which examines specific research on the detection and untangling of tangled commits. The Section \ref{subsec:natural-language-processing}, discusses the evolution of Natural Language Processing (NLP) techniques, from traditional methods to modern LLMs. The Section \ref{subsec:llms-in-software-engineering} provides a comprehensive review of the application of LLMs across diverse software engineering (SE) tasks. The Section \ref{subsec:explainable-artificial-intelligence} details approaches that enhance  interpretability and transparency in artificial intelligence (AI) models. Finally, \ref{subsec:positioning-our-work} positions the contributions of this study within the context of the related literature.

\subsection{PR-Issue Alignment}
\label{subsec:pr-issue-alignment}
In software development, a commit is ideally a self-contained change that addresses a single concern, such as fixing a bug or adding a new feature \citep{microsoft2024pullrequests}. This practice is fundamental for maintaining clear traceability and enabling effective collaboration. The precise alignment of a commit code changes with the requirements of its corresponding issue is a foundational goal for structured development workflows. However, empirical evidence shows that this ideal is often not achieved in practice, with developers frequently including unrelated changes that complicate the intended relationship between the commits and its corresponding issue \citep{herzig-msr-2013}.

\citet{herzig-msr-2013} were among the earliest researchers to systematically investigate the detrimental effects of Tangling commits. They define a tangling commit as one that combines unrelated changes into a single unit, thereby compromising the intended self-contained nature of a change. Through an empirical study of five open-source Java projects, they provided concrete evidence of the problem's prevalence, demonstrating that up to 20\% of bug-fixing commits were tangled. This research highlighted how such commits introduces substantial noise into software repository data, which can skew defect prediction models and lead to a considerable portion of files being incorrectly associated with bug reports. Their proposed heuristic-based untangling algorithm further demonstrated that up to 16.6\% of files may be incorrectly associated with bug reports due to noise introduced by tangling commits. 

 \citet{herbold2022fine} has shown the influence of tangled commits on the reliability and accuracy of software engineering analyses, particularly affecting the alignment between commits and their associated issues. Their study revealed that approximately 47\% of commits in open-source repositories involve task mixing, where unrelated changes are bundled within a single commit, thereby introducing noise that can adversely affect tasks such as defect prediction, and bug localization. While existing heuristics and clustering-based untangling tools can detect many of these tangled commits, around 20\% remain challenging to identify due to subtle interdependencies or complex code structures. These undetected cases can distort research outcomes, particularly when correlating code changes with issues or tracking software evolution over time. 

Recent work by \citet{bilsen} extended the alignment relationship between PR-issue pairs and proposed a new PR–issue alignment taxonomy comprising four distinct classes: Exact, where a PR fully addresses the associated issue without introducing irrelevant changes; Missing, where the PR fails to address the issue; Tangling, where unrelated changes are included; and Missing and Tangling, which combines both misalignment types. Analysis of a sampled subset of PRs from the Transformers repository\footnote{\url{https://github.com/huggingface/transformers/}} reveals that 68.04\% are labeled as Exact, 16.5\% as Missing, 13.4\% as Tangling, and 2.06\% as both Missing and Tangling. They also investigated the relationship between merge status and alignment, showing that 123 of 163 merged PRs (75.46\%) were classified as Exact compared to 9 of 31 closed unmerged PRs (29.03\%), suggesting a potential correlation between a PR's merged status and its likelihood of exact alignment. Furthermore, they examined how effectively LLMs identify PR-issue alignment inconsistencies with different prompt configurations.

\subsection{Tangling Commits}
\label{subsec:tangling-commits}

\citet{Untangling-fine-grained-code-changes} developed \textit{EpiceaUntangler}, focused on tangling commits, to automatically cluster changes into coherent, self-contained commits using fine-grained edit histories captured in real time from developers’ IDEs. The tool achieved a median untangling accuracy of 91\% (minimum 88\%), relying on features such as timing between changes, unrelated edits, and class-level modifications. While highly accurate, this approach required IDE instrumentation and developer interaction, limiting its applicability.

To address these limitations, \citet{flexeme} introduced \textit{Flexeme}, a language-agnostic graph-based method that employs a multiversion Program Dependency Graph ($\delta$-NFG) to track identifier evolution across versions. Although its untangling accuracy (81\%) was slightly lower than \textit{EpiceaUntangler}, \textit{Flexeme} improved automation and broadened applicability beyond IDE-bound solutions, highlighting the need for scalable approaches in complex repositories.

Building on this progression, \citet{untangling-composite-commits-by-attributed-graph-clustering} proposed \textit{ComUnt}, which framed commit untangling as an attributed graph clustering problem. By leveraging graph clustering techniques, \textit{ComUnt} achieved a 7.8\% performance improvement over earlier automated methods. However, like \textit{Flexeme}, its success was constrained by graph representation quality and the difficulty of tuning unsupervised clustering thresholds.

To overcome the shortcomings of unsupervised methods, \citet{utango} developed \textit{UTango}, a supervised learning approach that adapted agglomerative clustering to developer-specific grouping patterns. This strategy increased adaptability and contextual awareness but introduced new challenges, notably the dependence on high-quality labeled commit datasets, which are scarce and costly to obtain.

Recognizing that neither fully automated nor purely manual strategies are sufficient, \citet{ChangeBeadsThreader} presented \textit{ChangeBeadsThreader} as a human-in-the-loop solution. By allowing developers to refine clustering outcomes through splitting, merging, or adjusting commits, this approach balanced automation with necessary human oversight.

On the other hand, \citet{are-you-committing-tangled-changes} proposed a preventive approach designed to encourage better commit practices during development. Their method analyzes past commits to derive single-task templates and alerts developers in real time when new commits resemble multi-task patterns. This proactive strategy promotes atomic commits and improves the alignment of changes with tracked issues, thereby reducing the downstream need for untangling.

\subsection{Natural Language Processing}
\label{subsec:natural-language-processing}
Text classification has long served as a foundational task in NLP, evolving from traditional statistical models to modern large-scale pre-trained transformers. Early approaches relied on algorithms such as Naïve Bayes, logistic regression, and support vector machines (SVMs), typically applied over bag-of-words or TF-IDF representations \citep{joachims1998text, sebastiani2002machine}. These models, although interpretable and computationally efficient, were limited in their ability to capture semantic and syntactic structures. In software engineering, they were primarily employed for tasks such as automated bug report classification \citep{anvik2006automated} and sentiment analysis of developer comments \citep{sentiment-analysis-software}.

The introduction of word embeddings, particularly Word2Vec \citep{Word2Vec} and GloVe \citep{glove}, marked a significant turning point by providing distributed vector representations that captured semantic relationships among words. These embeddings enabled the use of neural architectures such as convolutional neural networks (CNNs) \citep{kim2014convolutionalneuralnetworkssentence} and recurrent neural networks (RNNs), including long short-term memory (LSTM) networks \citep{LSTM}, which allowed the modeling of deeper contextual dependencies.

A major paradigm shift occurred with the introduction of transformer-based architectures. Models such as Bidirectional Encoder Representations from Transformers (BERT) \citep{devlin2018bert} demonstrated the potential of large-scale pre-training followed by fine-tuning for downstream tasks. BERT’s masked language modeling (MLM) objective facilitated bidirectional context learning, achieving state-of-the-art results on multiple NLP benchmarks such as GLUE \citep{wang2019gluemultitaskbenchmarkanalysis}. In software engineering contexts, encoder-based variants like CodeBERT \citep{feng2020codebert} and GraphCodeBERT \citep{guo2021graphcodebertpretrainingcoderepresentations} were fine-tuned for code-related tasks including function naming, clone detection, and issue classification. Expanding this line of research, unified frameworks such as UniXcoder \citep{guo2022unixcoder} integrated natural language and code representations to enhance multi-modal understanding.

Subsequent developments extended transformers into encoder-decoder and decoder-only architectures, including T5 \citep{raffel2020exploring} and CodeLLaMA \citep{roziere2023code}, respectively, which combine task generalization with generative capabilities. Such models, trained on both natural language and large-scale code corpora, provide a flexible foundation for downstream fine-tuning in software engineering tasks.

LLMs such as GPT-3, GPT-4, and LLaMA \citep{brown2020language, touvron2023llama} represent a distinct evolution of pre-trained transformers. Unlike earlier models that relied heavily on supervised fine-tuning, LLMs excel at zero-shot and few-shot learning via prompt-based classification \citep{santana2025prompting}. This paradigm enables rapid adaptation to new tasks without retraining, making them suitable for general purpose tasks. However, challenges persist in applying LLMs to fine-grained classification problems in SE, particularly with noisy or lengthy inputs such as code diffs and issue discussions. Prompt-based approaches, while promising, are sensitive to input design, lack robustness in domain-specific scenarios, and often underperform on tasks requiring precise alignment between natural language and code \citep{zhu2024promptrobust}.

These limitations highlight the need for complementary adaptation techniques, such as fine-tuning and parameter-efficient learning, which allow pre-trained LLMs to be tailored to software engineering domains. Current research trends increasingly emphasize hybrid approaches, where pre-trained LLMs are fine-tuned or adapted to SE domains, thereby balancing generalization capacity with domain specificity \citep{hou2024large}.

\subsection{LLMs in SE Tasks}
\label{subsec:llms-in-software-engineering}
LLMs have been applied to automate PR title and description generation, addressing incomplete or missing PR information. Approaches such as \textit{AutoPRTitle} leverage sequence-to-sequence models like BART to generate concise PR titles from PR descriptions, commit messages, and associated issue titles, achieving significant improvements over baseline methods in both ROUGE metrics and human evaluation \citep{9978252, 8952330}. These models reduce the burden on developers and enhance the clarity and consistency of PR metadata.

LLMs are also utilized for PR-issue analysis, including duplicate PR detection and issue linkage prediction. Techniques rely on textual similarity between PRs and issues, integrating features from titles, descriptions, labels, and comments \citep{10002372, 9978252}. These studies demonstrate that LLM-based representations improve the identification of semantically similar or duplicate PRs, supporting tasks such as issue tracking, feature location, and defect localization.
Beyond PR-issue analysis, LLMs also facilitate automated code generation from code review comments and diffs. For example, at Google, LLMs assist in generating suggested edits in response to reviewer comments, streamlining the review process and reducing developer workload \citep{10.1145/3131704.3131725}. These models interpret natural language instructions in code reviews and produce syntactically correct code modifications, demonstrating effective integration of LLMs into real-world software engineering workflows.

Automatic commit message generation is another key application. LLMs, including GPT and LLaMA variants, have been evaluated against traditional generation-based and retrieval-based models, demonstrating superior quality in human evaluations despite mixed results on BLEU and ROUGE metrics \citep{lopes2024commit, 10589767}. These models provide consistent and semantically accurate descriptions of code changes, improving codebase understanding and collaboration efficiency.
LLMs have been explored for classifying PR-issue alignment according to taxonomies such as Exact, Missing, Tangling, and Missing \& Tangling \citep{bilsen}. By encoding both PR and issue content, LLMs can capture nuanced semantic relationships, supporting automated assessment of PR-issue consistency and facilitating large-scale repository analysis.

In addition to direct task applications, LLMs have also been utilized for data augmentation in software engineering contexts. While general NLP studies have systematically explored augmentation strategies such as paraphrasing, back-translation, and synthetic data generation to expand training corpora \citep{feng2021survey}, foundational models like PLBART \citep{ahmad2021unified} have demonstrated how joint pre-training on code and natural language can enable generative capabilities later leveraged for producing synthetic artifacts such as commit messages, issue reports, and code comments. These generative techniques can contribute to expanding limited labeled datasets while maintaining semantic alignment between code and accompanying natural language artifacts.

\subsection{Explainable AI}
\label{subsec:explainable-artificial-intelligence}
Explainable AI methods aim to make model predictions understandable by quantifying the contribution of input features to outputs. Among these, SHAP is a model-agnostic framework that assigns each feature an importance value for a particular prediction based on cooperative game theory \citep{lundberg2017unified}. By decomposing the prediction into additive contributions of input features, SHAP allows researchers to quantify the effect of each feature on model outputs.

Local Interpretable Model-agnostic Explanations (LIME) provides local explanations for individual predictions by approximating the original model with an interpretable surrogate model in the vicinity of a given input \citep{ribeiro2016model}. LIME generates feature-level contributions that clarify how input perturbations affect predictions. More broadly, within the field of AI, it has been used to interpret outputs of complex models across diverse application domains, thereby enhancing the transparency of model decision-making processes.


Integrated Gradients (IG) is a gradient-based method for attributing the prediction of deep neural networks to input features \citep{sundararajan2017axiomatic}. By integrating gradients along a path from a baseline input to the actual input, IG provides a theoretically grounded measure of feature importance. In software engineering applications, IG can be applied to understand how LLMs and transformer models weigh code tokens or textual components when generating outputs, offering insight into the internal mechanisms behind model predictions.

\subsection{Positioning our Work}
\label{subsec:positioning-our-work}

Following \citet{herzig-msr-2013}’s introduction of tangling commits, most studies focused on this tangling commits. Later research expanded to include methods for untangling tangled commits, preventing tangling commits, and analyzing its resulting impacts \citep{herzig-msr-2013,are-you-committing-tangled-changes,Untangling-fine-grained-code-changes,ChangeBeadsThreader,flexeme,untangling-composite-commits-by-attributed-graph-clustering,utango,herbold2022fine}. These approaches included both heuristic methods, preventive approaches, and graph-based clustering techniques to partition or refine commit histories. Although effective in identifying tangled changes, these studies largely concentrated on a single dimension of commits. The study by \citet{bilsen} marked a notable advance by extending the alignment concept to PR–issue pairs and introducing a new taxonomy with four categories: Exact, Missing, Tangling, and Missing and Tangling. As shown in Table~\ref{tab:pr_issue_studies}, their evaluation combined effects analysis and detection of PR-issue misalignments with LLM prompting. However, the reliance on prompt engineering limited the extent to which model capabilities were fully utilized, as fine-tuning was not explored.

\begin{table}[h]
\centering
\scriptsize
\caption{Comparison of PR-issue Alignment Studies}
\begin{tabular}{@{}p{2.4cm}p{1cm}p{1.5cm}p{6cm}@{}}
\toprule
\textbf{Study} & \textbf{Research Focus} & \textbf{Taxonomy} & \textbf{Method} \\
\midrule
\citet{herzig-msr-2013} & E, U & T & Heuristic analysis of bug-fixing commits to quantify prevalence and impact of tangled changes \\
\midrule
\citet{are-you-committing-tangled-changes} & P & T & Real-time commit template checking to guide developers toward atomic commits \\
\midrule
\citet{Untangling-fine-grained-code-changes} & D, U & T & Fine-grained IDE-based edit history clustering to partition changes into self-contained commits \\
\midrule
\citet{ChangeBeadsThreader} & D, U & T & Human-in-the-loop refinement of clustered commits via interactive splitting, merging, and adjustment \\
\midrule
\citet{flexeme} & D, U & T & Graph-based identifier evolution tracking across versions using a multiversion Program Dependency Graph (\(\delta\)-NFG) \\
\midrule
\citet{untangling-composite-commits-by-attributed-graph-clustering} & D, U & T & Attributed graph clustering applied to commit graphs for systematic untangling \\
\midrule
\citet{utango} & D, U & T & Supervised agglomerative clustering adapted to developer-specific grouping patterns \\
\midrule
\citet{herbold2022fine} & E & T & Repository mining with statistical analysis to assess task-mixed commits \\
\midrule
\citet{bilsen} & D & Ex, M, T, MT & Manual PR labeling combined with LLM-based prompting to classify PR–issue alignment \\
\rowcolor{blue!15} Our Work & D & Ex, M, T, MT & Fine-tuning of open-source and proprietary LLMs to detect PR–issue alignment across multiple categories, with interpretability analysis to identify the most influential PR–issue fields. \\
\bottomrule
\end{tabular}
\footnotesize  \\
\textbf{Research Focus abbreviations:} E = Effects, P = Preventive,\\ D = Detection, U = Untangling\\
\textbf{PR-Issue Taxonomy abbreviations:} Ex = Exact, M = Missing, T = Tangling, MT = Missing and Tangling
\label{tab:pr_issue_studies}
\end{table}

Our work, highlighted in Table~\ref{tab:pr_issue_studies}, addresses this gap. In contrast to earlier heuristic or untangling-focused approaches, and beyond \citet{bilsen}’s prompting approach, we adopt a fine-tuning strategy applied to both closed-source and open-source LLMs. By adapting models specifically for PR–issue alignment, our methodology leverages adaptation rather than ad-hoc prompting. Furthermore, interpretability analysis helps us to understand which PR–issue fields influence fine-tuned LLMs decisions, providing additional insights into the LLMs decision process.

%% file: methodology.tex
\section{Methodology}
\label{sec:methodology}

\begin{figure*}[h]
  \centering
  \includegraphics[width=\textwidth]{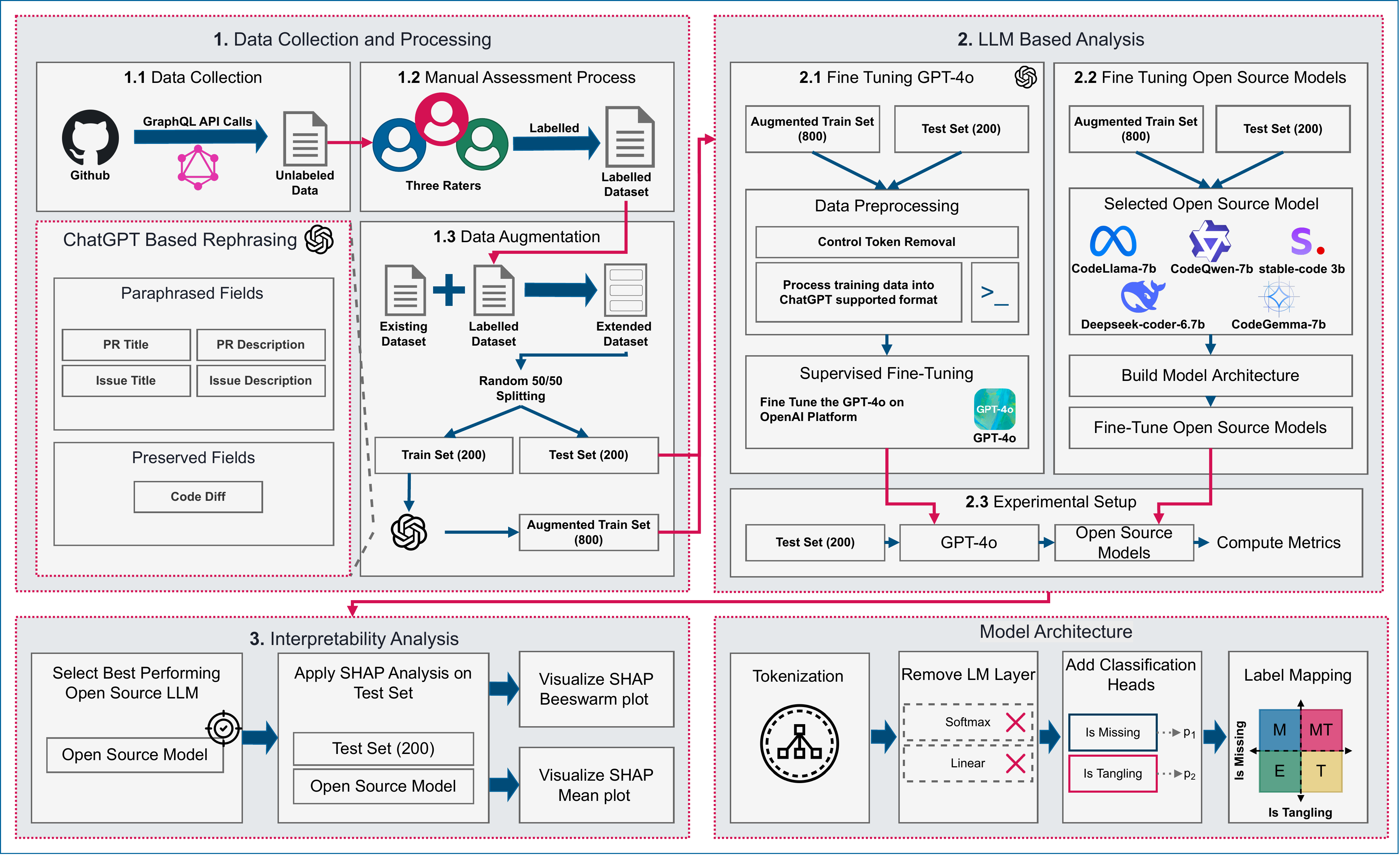}
\caption{Overview of Methodology}
\label{fig:methododogy}
\end{figure*}

Our methodology is designed to address all of the RQs. To address RQ1 and RQ2, we first extended a existing dataset from \citet{bilsen} and then used data augmentation to resolve a class imbalance as shown in Step 1 of Fig. \ref{fig:methododogy}. We subsequently fine-tuned the GPT-4o, a state-of-the-art model, and open-source LLMs CodeLLaMA-7B, CodeQwen1.5-7B, StableCode-3B, CodeGemma-7B, and Deepseek-Coder-6.7B.

We selected these LLMs because they have fewer than 10 billion parameters, ensuring the feasibility of fine-tuning on a single 40GB GPU while preserving architectural diversity. These models represent a range of code-specialized transformer architectures capable of understanding both natural language and programming semantics. Each LLM was pre-trained on large-scale source-code corpora and natural-language documentation, enabling them to capture hybrid relationships between textual and structural information in PR–issue pairs. Specifically, CodeLLaMA and CodeGemma extend the LLaMA and Gemma families with enhanced code reasoning capabilities; CodeQwen1.5 introduces bilingual code–text alignment beneficial for mixed-language repositories; StableCode emphasizes lightweight design for resource-constrained environments; and Deepseek-Coder provides extensive exposure to GitHub-scale repositories. Collectively, these LLMs also balance domain specialization, scalability, and compatibility with parameter-efficient fine-tuning (PeFT) approaches, making them well-suited for the PR–issue alignment classification task.

We applied supervised fine-tuning on both open-source and closed-source LLMs. For the closed-source LLM, GPT-4o, we employed the instruction-tuning as a fine-tuning method, which was the most suitable approach offered by OpenAI. For the open-source LLMs, text-generation head was removed, two new classification heads were added, and only these classifiers were fine-tuned. This was conducted to evaluate the performance improvement from fine-tuning, compare it with the previous approach \citep{bilsen} as shown in Step 2 of Fig \ref{fig:methododogy}. To maintain comparability with prior research \citep{bilsen}, the performance of the fine-tuned LLMs on the PR-issue alignment classification task is evaluated through accuracy, F1-score (micro, macro, and weighted), precision, recall, and specificity  metrics. 

To address RQ3, we applied SHAP analysis to the open-source LLM with the highest accuracy—CodeLlama-7B—as illustrated in Step 3 of Fig. \ref{fig:methododogy}.

The remainder of this section outlines the various components of our methodology. Section~\ref{subsec:data-collection-and-processing} describes the data collection, manual assessment, and augmentation processes. Subsequently, Section~\ref{subsec:llm-based-analysis} outlines the model-specific fine-tuning process, including its architecture, training configurations, and experimental pipeline. Finally, the interpretability analysis, which utilizes SHAP, details how different dataset fields influence on PR-issue classification predictions.

\subsection{Data Collection and Processing}
\label{subsec:data-collection-and-processing}
The initial dataset used for this study was sourced from \citet{bilsen}, containing 163 merged and 31 unmerged PR–issue pairs, distributed across four classes: Exact 132 (68.4\%), Tangling 26 (13.4\%), Missing 32 (16.5\%), and Missing and Tangling 4 (2.6\%). Then we extended this dataset to a two times  Cochran sample size \citep{cochran1977sampling}, yielding representative training and test sets. The additional instances required were selected from the Transformers repository, a prominent NLP library written in Python, which includes over 2,000 closed PR–issue pairs, ensuring consistency with the existing dataset.

Following manual labeling and merging with the existing dataset, we observed a class imbalance, with the Exact class being predominant. The final distribution of PR–issue alignment labels was as follows: Exact 267 (66.75\%), Missing 59 (14.75\%), Tangling 56 (14\%), and Missing and Tangling 18 (4.50\%). To address this imbalance, we applied data augmentation exclusively to the training split, equalizing class sizes to provide a more balanced dataset for model learning. This decision was made to mitigate the model’s tendency to overfit the majority class and to enable a fairer evaluation of its capacity to distinguish between the rarer alignment types (Missing, Tangling, Missing and Tangling), thereby promoting a fair and stable fine-tuning process across all categories.

\input{methodology-data-collection-and-dataset-preparation}

\input{methodology-manual-assessment-process}

\input{methodology-data-augmentation}

\subsection{LLM-Based Analysis}
\label{subsec:llm-based-analysis}
To analyze the effectiveness of fine-tuned LLMs, we utilized supervised fine-tuning for both proprietary and open-source models. GPT-4o was fine-tuned by instrunction-tuning method using OpenAI’s supervised fine-tuning API. For the open-source models (CodeLLama-7B, CodeQwen1.5-7B, StableCode-3B, CodeGemma-7B, and Deepseek-Coder-6.7B), we removed the text-generation head, added two classification heads, and fine-tuned only these heads to adapt to hardware constraints. To ensure comparability with \citet{bilsen}, we standardized the input configuration across all models to include the issue title, issue description, PR title, PR description, and the code diff. This alignment of input fields enables a consistent evaluation framework between our experiments and prior research.

\input{methodology-fine-tuning-chatgpt}
\input{methodology-fine-tuning-open-source-models}
\input{methodology-llm-based-analysis-experimental-setup}
\input{methodology-interpretability-analysis}

%% file: methodology-data-collection-and-dataset-preparation.tex
\subsubsection{Data Collection}
\label{subsubsec:data-collection}

Manually evaluating all PR-issue pairs to construct a ground-truth dataset is an infeasible task due to the massive volume of data in large-scale software repositories. To address this challenge and ensure the statistical representativeness of our findings, we utilized a random stratified sampling approach \citep{baltes2022sampling}. This method allows us to select a statistically significant subset of the data that maintains the same distribution of key characteristics as the entire set of PR-issue pairs within the repository, thereby ensuring accuracy and generalizability to our study. In our case, the only prior data available was the merge status of PRs in the transformers repository, which was the sole basis for our stratification. The required sample size was calculated using Cochran's formula \citep{cochran1977sampling}, and PR-issue pairs fetched from the repository and prepared for the labeling process. \\


\textit{Sample Selection:} As of July 2, 2025, the Transformers repository contained 2,831 merged and 563 unmerged PRs explicitly linked to issues. Using a $95\%$ confidence level ($z = 1.96$ \citet{moore2009introduction}), a $5\%$ margin of error, and merged PR proportion of $p = 0.834$, the Cochran sample size formula indicated a required sample size of 200  (167 merged and 33 unmerged) instances are required to represent the behavior of the original dataset accurately. We doubled this number and obtained a total of 400 labeled instances. Our initial dataset had 163 merged and 31 unmerged PR-issue pairs. To achieve the target sample size while preserving the original class distribution, we randomly sampled an additional 171 merged and 35 unmerged PR-issue pairs that were not already present in the existing dataset. \\


\textit{Data Extraction:} The data extraction process was conducted in two primary steps using the GitHub GraphQL API:
\begin{enumerate}
    \item \textit{Fetching URLs and Merge Statuses:} A custom Python script was used to fetch the URLs and merge statuses of closed PR-issue pairs that were not included in our existing dataset. These URLs were then used directly in the manual annotation process.
    \item \textit{Retrieving Detailed Metadata:} After the manual assessment was complete, a second extraction was performed to fetch the details for each labeled PR-issue pair. Fetched fields included the PR titles, PR descriptions, issue titles, issue descriptions, and code diffs. This data was then merged with the manual labeling results to create our final, comprehensive dataset for model training and evaluation.
\end{enumerate}

%% file: methodology-manual-assessment-process.tex
\subsubsection{Manual Assessment Process}
\label{subsubsec:manual-assessment-process}


The manual labeling of PR–issue pairs was conducted in two stages to ensure both accuracy and consistency. In the first stage, two raters with Python experience were involved: one with over 3 years, and the other with over 7 years of experience. Each rater independently evaluated the PR-issue pairs and assigned them to one of four classes: Exact, Tangling, Missing, and Missing and Tangling. Prior to labeling, the raters consulted with annotators from the previous study \citep{bilsen} to obtain additional context and align with the established evaluation protocol. Each PR--issue pair was assessed using the following fields:
\begin{enumerate}
    \item \textit{PR Diff:} Evaluated the relevance and appropriateness of code changes in relation to the associated issue.
    \item \textit{PR and Issue Titles and Descriptions:} Examined PR and issue titles and descriptions to assess the relevance and alignment between the linked PR and its associated issue.
    \item \textit{PR and Issue Discussions:} Examined both PR and issue discussion threads to assess rationale, context, and any reasoning behind acceptance or rejection decisions, including reviewer comments and suggested modifications.
\end{enumerate}


In the second stage, a third rater with over 20 years of experience was involved to resolve conflicts from the first stage and assign the final labels. This rater reviewed all discrepancies, provided final label assignments, and added detailed notes and comments to maintain transparency and traceability, adhering to the evaluation protocol to ensure reliable labeling.

Following the two-stage evaluation, Cohen's Kappa score ($\kappa = 0.575$) and Inter-Rater Reliability ($IRR = 0.791$) were calculated to assess consistency. These metrics indicate moderate agreement between raters and a dependable labeling process. The resulting labeled dataset consists of 137 ($66.50\%$) Exact , 38 ($18.44\%$) Missing, 22 ($10.67\%$) Tangling, and 9 ($4.36\%$) Missing and Tangling instances.

%% file: methodology-data-augmentation.tex
\subsubsection{Data Augmentation and Dataset Preparation}
\label{subsubsec:data-augmentation-and-dataset-preparation}


Following the manual labeling process, the labeled dataset was merged with the existing dataset. The extended dataset was then split into train and test sets using a 50\%-50\% random stratified sampling based on merge status, ensuring the preservation of the actual merge status distribution across both sets. The resulting training set contained 131 (65.5\%) Exact, 30 (15\%) Tangling, 28 (14\%) Missing, and 11 (5.5\%) Missing and Tangling PR–issue pairs, showing a significant class imbalance. To address the class imbalance and standardize class representation, the training set was augmented using LLMs to generate additional examples until each class contained 200 instances, resulting in an 80\%-20\% ratio between training and test sets. This augmentation strategy not only ensured balanced class representation in the training set but also expanded the limited dataset, thereby increasing the diversity and availability of training examples, while preserving the original samples in the test set.

During the augmentation process, the textual fields of each PR and issue pair, including the PR title, PR description, issue title, and issue description systematically rephrased using LLMs, while the code diffs were preserved to maintain the fidelity of the underlying implementation. Each field was rephrased independently to ensure both clarity and coherence. To preserve the logical relationship between titles and descriptions within each PR–issue pair, each field was provided to the LLM within a structured prompt format that included the necessary context, with the system prompt and the corresponding user prompt for each field  detailed in Table \ref{tab:pr_issue_rephrase_prompt}.

\begin{table}[h]
    \centering
    \renewcommand{\arraystretch}{1.2}
    \setlength{\tabcolsep}{4pt}
    \caption{PR/Issue Rephrasing Prompts for LLM}
    \begin{tabular}{|p{0.25\linewidth}|p{0.7\linewidth}|}
        \hline
        \multicolumn{2}{|l|}{\textbf{System Prompt}} \\
        \hline
        \multicolumn{2}{|p{0.95\linewidth}|}{You are an expert software engineering assistant specializing in pull request and issue analysis. Rephrase the given PR or issue field while preserving its technical meaning, alignment intent, context, and length of the original text. Return only the rephrased text with natural language fluency, without explanations or additional commentary. If the input is empty or null, return an empty string.} \\
        \hline
         \textbf{Field} & \textbf{User Prompts} \\
        \hline
        PR Title & You are given a Pull Request title and description. Rephrase only the title, ensuring consistency with the description and preserving its technical context. PR title: \textcolor{springernatureblue}{\textbf{\{PR title\}}} PR description: \textcolor{springernatureblue}{\textbf{\{PR description\}}}. \\
        \hline
        PR Description & You are given a Pull Request title and description. Rephrase only the description, ensuring consistency with the title and preserving its technical context. PR title: \textcolor{springernatureblue}{\textbf{\{PR title\}}} PR description: \textcolor{springernatureblue}{\textbf{\{PR description\}}}. \\
        \hline
        Issue Title & You are given an Issue title and description. Rephrase only the title, ensuring consistency with the description and preserving its technical context. Issue title: \textcolor{springernatureblue}{\textbf{\{issue title\}}} Issue description: \textcolor{springernatureblue}{\textbf{\{issue description\}}}. \\
        \hline
        Issue Description & You are given an Issue title and description. Rephrase only the description, ensuring consistency with the title and preserving its technical context. Issue title: \textcolor{springernatureblue}{\textbf{\{issue title\}}} Issue description: \textcolor{springernatureblue}{\textbf{\{issue description\}}}. \\
        \hline
    \end{tabular}
    \label{tab:pr_issue_rephrase_prompt}
\end{table}

For data augmentation, OpenAI ChatGPT-4.1 was utilized because it is specialized in text generation, which suits our text rephrasing needs. During augmentation, the OpenAI Batch API was used to efficiently process multiple PR and issue entries in parallel. The model temperature was set to 0.2, optimizing the balance between reproducibility and creative variation in the generated fields \citep{alshahwan2024automated, guilherme2023initial}.

%% file: methodology-fine-tuning-chatgpt.tex
\subsubsection{Fine-Tuning GPT-4o}
\label{subsubsec:fine-tuning-chat-gpt}

\begin{figure*}[h]
  \centering
  \includegraphics[width=\textwidth]{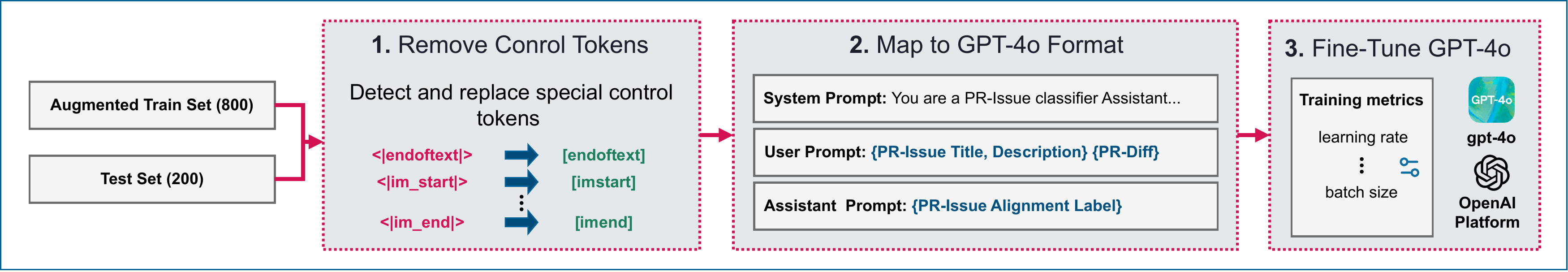}
\caption{Fine-tuning GPT-4o workflow}
\label{fig:fine_tuning_chatgpt}
\end{figure*}

The PR-issue alignment classification task modeled as a text generation task, where the model generates the class label directly. The instruction tuning process began with the preparation of the training and test sets by removing special control tokens such as \textnormal{\textless{}\textbar{}im\_start\textbar{}\textgreater{}}, 
\textnormal{\textless{}\textbar{}im\_end\textbar{}\textgreater{}} and \textnormal{\textless{}\textbar{}endoftext\textbar{}\textgreater{}}.These tokens were replaced with representative placeholders to ensure compatibility with GPT-4o’s training requirements while maintaining the semantic integrity of the input, as illustrated in Step 1 of Fig.~\ref{fig:fine_tuning_chatgpt}, ensuring compatibility with GPT-4o's training requirements.

Subsequently, the datasets were mapped into the chat completion format required for fine-tuning GPT-4o, as illustrated in Step 2 of Fig.~\ref{fig:fine_tuning_chatgpt}. This format consisted of three components: a system prompt, a user prompt, and an assistant prompt. The system prompt provided a instruction that described the classification task and outlined the possible labels, namely Exact, Missing, Tangling, and Missing and Tangling. The user prompt included structured information about the PR-Issue pair, including the issue title, issue description, PR title, PR description, and code diffs. The assistant response corresponded to the ground truth alignment label. We used the same prompt as proposed by \citet{bilsen}, which demonstrated the highest accuracy. The complete set of prompts, including the system, user, and assistant roles, is presented in Table \ref{tab:pr_issue_alignment_prompt}.

The curated dataset was then used to fine-tune the gpt-4o-2024-08-06 model by instruction tuning on the OpenAI platform, as shown in Step 3 of Fig. ~\ref{fig:fine_tuning_chatgpt}. Fine-tuning was performed with: 3 epochs, a batch size of 16, and a learning rate of $3\times10^{-5}$ to optimize the model for PR-Issue alignment classification.

\begin{table}[h]
    \centering
    \renewcommand{\arraystretch}{1.2}
    \setlength{\tabcolsep}{4pt}
    \caption{GPT-4o Prompts for Fine-Tuning}
    \begin{tabular}{|p{0.1\linewidth}|p{0.85\linewidth}|}
        \hline
        \textbf{Role} & \textbf{Prompt} \\
        \hline
        System & You are a rigorous code review assistant. Your task is to classify the alignment between Pull Requests (PRs) and the specified issues based on how well they align with their corresponding issues. Note that issues may describe a bug, request a feature, or provide general guidance, and do not always include specific implementation details. If implementation details are provided in the issue, the PR should follow them closely. If no specific implementation is given in the issue, consider the PR ``Exact'' if it fully addresses the issue without any missing or unrelated changes. Use these specific labels for classification: \textbf{Exact:} The PR fully implements all parts of the issue, without adding unrelated changes. \textbf{Missing:} The PR does not include one or more parts specified in the issue. Even if an issue has multiple parts to be handled by different PRs, each PR must independently fully solve one part of the issue. \textbf{Tangling:} The PR includes extra, unneeded changes not requested in the issue. This can include unrelated refactoring, changes in other parts of the codebase not required for the solution, or additional modifications that do not contribute directly to resolving the issue. \textbf{Missing and Tangling:} The PR is both incomplete and contains unrelated changes. It is missing parts of the issue requirements and includes unrelated changes (such as refactoring or unrelated modifications) that do not contribute to the issue’s solution. Only respond with one of the following labels: Exact, Missing, Tangling, Missing and Tangling. Do not provide any additional explanation or text. \\
        \hline
        User & Analyze the following PR-Issue pair and classify the alignment between them. Only respond with one of the following labels: Exact, Missing, Tangling, Missing and Tangling. Do not provide any additional explanation or text.
        Issue Title: \textcolor{springernatureblue}{\textbf{\{Issue Title\}}} 
        Issue Description: \textcolor{springernatureblue}{\textbf{\{Issue Description\}}} 
        PR Title: \textcolor{springernatureblue}{\textbf{\{PR Title\}}}
        PR Description: \textcolor{springernatureblue}{\textbf{\{PR Description\}}}
        Code Changes: \textcolor{springernatureblue}{\textbf{\{PR Diff\}}}
        Based on the provided information, classify the PR-Issue alignment. \\
        \hline
        Assistant & \textcolor{springernatureblue}{\textbf{\{PR-Issue Alignment Label\}}} \\
        \hline
    \end{tabular}
    \label{tab:pr_issue_alignment_prompt}
\end{table}

%% file: methodology-fine-tuning-open-source-models.tex
\subsubsection{Fine-Tuning Open Source LLMs}
\label{subsubsec:fine-tuning-open-source-models} 


We modeled the PR–issue classification task as a dual binary-classification problem in the context of fine-tuning open-source LLMs. Specifically, two independent binary classifiers were trained: one to predict whether a PR–issue pair exhibits tangling, and another to predict whether it exhibits missing alignment characteristics. The outputs of these classifiers were then combined to derive the four alignment categories—Exact, Missing, Tangling, and Missing and Tangling—ensuring that the fine-tuned models captured both types of misalignment while remaining consistent with the PR–issue alignment framework.

\begin{figure*}[h]
  \centering
  \includegraphics[width=\textwidth]{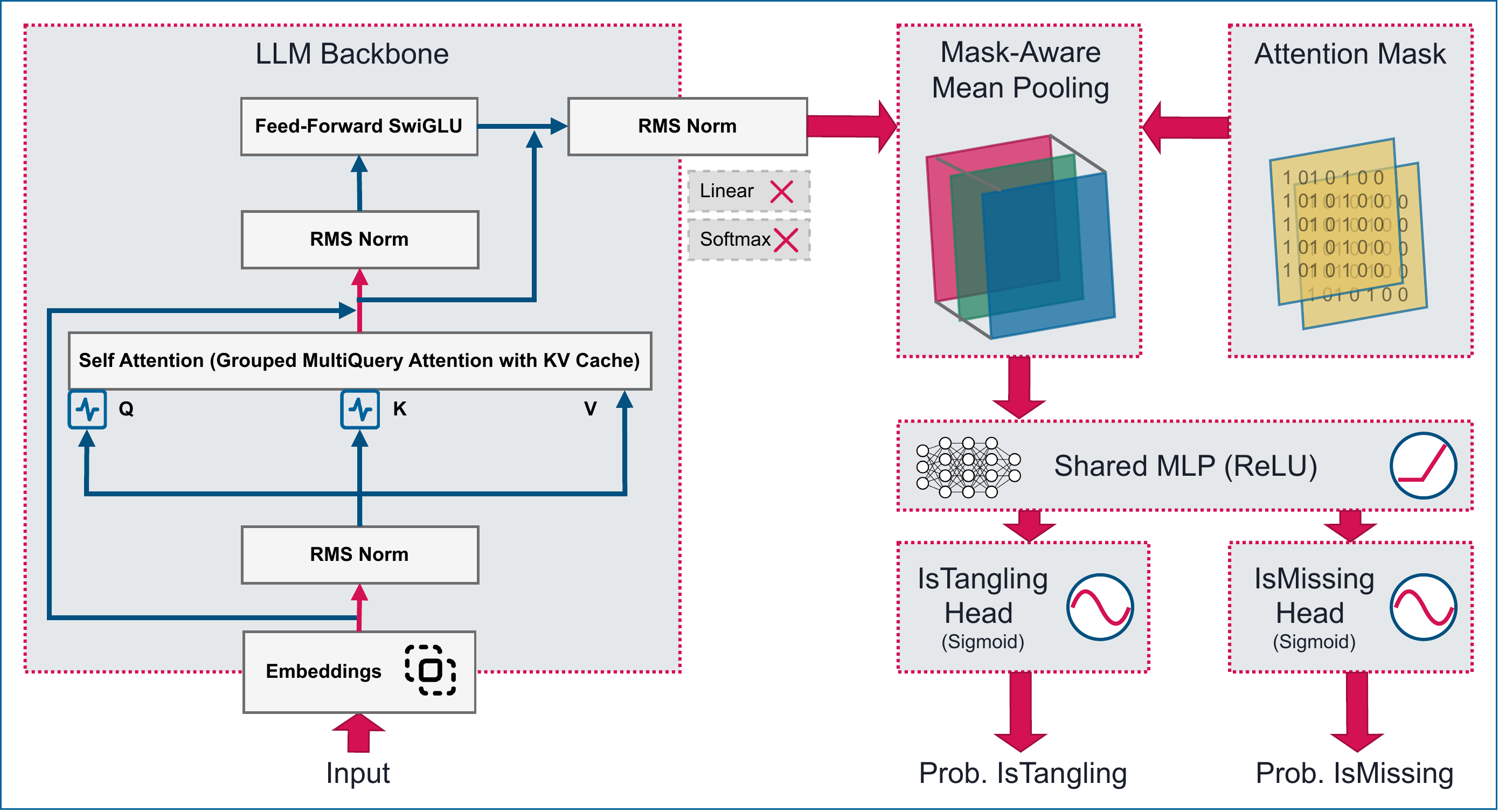}
\caption{Fine-tuned LLM Architecture For CodeLLaMA}
\label{fig:fine_tuning_open_source_llm}
\end{figure*}


\textit{Model Architecture:} To adapt the open-source LLMs to our PR–issue classification formulation, we first serialize each PR–issue pair into the prompt format (Table~\ref{tab:open-source-llm-prompt}) and tokenize it with the model’s native tokenizer, obtaining input IDs and an attention mask. The pretrained language modeling head (linear and softmax layers) is then removed and replaced with an attention-mask–aware mean pooling layer that aggregates token embeddings only over non-padding tokens indicated by the mask, yielding a fixed-length sequence representation. This pooled vector is fed into a shared multi-layer perceptron (MLP) comprising a linear projection, ReLU activation, and dropout regularization, followed by two independent classification heads. Each head outputs a single logit—one predicting whether a PR–issue pair exhibits \textit{missing} alignment and the other predicting \textit{tangling} alignment—which are converted to probabilities via sigmoid activations. This lightweight architecture, illustrated in Fig.~\ref{fig:fine_tuning_open_source_llm}, effectively captures PR–issue alignment signals while preserving computational efficiency and compatibility with our supervised fine-tuning pipeline.

To prepare the input for the open-source LLMs, we first determined the input format as shown in Table \ref{tab:open-source-llm-prompt}, which incorporated the PR–issue title, description, and code diffs. We then applied the appropriate tokenizer specific to each LLM and set the maximum context window to 8,192 tokens to maintain balance between capturing extended contextual coverage and maintaining computational efficiency under GPU memory constraints. When the concatenated input exceeded this length, excess tokens were truncated from the right side. This ensured that all inputs remained within the context constraints while retaining the most semantically relevant portions of the PR–issue pair, which are both titles, both descriptions, and the initial sections of code diffs, preserving contextual integrity for accurate classification. \\

\begin{table}[htbp]
\centering
\renewcommand{\arraystretch}{1.3}
\caption{Input prompt format for open-source LLMs.}
\begin{tabular}{|p{0.9\linewidth}|}
\hline
\textbf{Open-Source LLM Input Prompt Format} \\
\hline
\#\#\# ISSUE TITLE: \textcolor{springernatureblue}{\textbf{\{issue\_title\}}} \\
\#\#\# ISSUE BODY: \textcolor{springernatureblue}{\textbf{\{issue\_body\}}} \\ 
\#\#\# PR TITLE: \textcolor{springernatureblue}{\textbf{\{pr\_title\}}} \\
\#\#\# PR BODY: \textcolor{springernatureblue}{\textbf{\{pr\_body\}}} \\
\#\#\# CODE DIFF: \textcolor{springernatureblue}{\textbf{\{code\_diff\}}} \\
\hline
\end{tabular}
\label{tab:open-source-llm-prompt}
\end{table}


Finally, to complete the mapping from model outputs to the PR–issue alignment taxonomy, we translated the probabilistic predictions from the \textit{isMissing} and \textit{isTangling} heads into categorical alignment labels. For conciseness, we denote the output of the \textit{isMissing} head as $M$ and the output of the \textit{isTangling} head as $T$. As formalized in Equation \ref{eq:alignment}, each PR–issue pair is classified as Exact, Tangling, Missing, or Missing and Tangling based on thresholded outputs from the respective sigmoid activations. This mapping ensures that the binary predictions produced by the model are aligned with the four-class PR-issue alignment classes.

\begin{equation} \label{eq:alignment}
\text{Alignment}(M, T) =
\begin{cases} 
\text{Exact} & \text{if } M < 0.5 \text{ and } T < 0.5, \\
\text{Tangling} & \text{if } M < 0.5 \text{ and } T \geq 0.5, \\
\text{Missing} & \text{if } M \geq 0.5 \text{ and } T < 0.5, \\
\text{Missing and Tangling} & \text{if } M \geq 0.5 \text{ and } T \geq 0.5,
\end{cases}
\end{equation}

\noindent\textit{Fine-Tuning Process}: 
The fine-tuning of the open-source LLMs was conducted with a focus on efficiency and practicality while maintaining model performance. Training was executed on an NVIDIA A100 GPU with 40 GB of memory, with all parameters of the LLMs frozen, such that only the shared MLP and the two classification heads were updated. The mask-aware pooling operation introduced for sequence embeddings did not contain any trainable parameters, which further simplified the optimization process and reduced computational overhead.

Freezing the backbone provided several advantages. Firstly, it eliminated the need to backpropagate gradients through billions of parameters, significantly reducing computation time. Secondly, frozen layers did not require gradient storage, conserving GPU memory. Finally, the reduced number of trainable parameters allowed the classifier to converge in fewer epochs, optimizing both time and resources while maintaining effective fitting to the training set.

To further optimize training, we employed 8-bit quantization for the backbone, allowing larger models to fit within the GPU memory constraints, and half-precision (bfloat16) arithmetic, which reduced memory footprint and accelerated computation without compromising numerical stability. The effective batch size was fixed at 16 for all experiments. When GPU memory was limited, we utilized gradient accumulation to simulate larger batches by accumulating gradients over multiple steps before performing an update, ensuring consistency across experiments.

Models were trained for $10$ epochs using the AdamW optimizer (\texttt{adamw\_torch\_fused}) with a learning rate of $3\times10^{-5}$, $\beta_1=0.9$, $\beta_2=0.999$, $\epsilon=1\times10^{-8}$, no weight decay, and a gradient clipping threshold of $0.9$, together with a linear learning rate scheduler (without warm-up steps) and a dropout rate of $0.1$ in the MLP. The loss function was defined as the average of two binary cross-entropy losses, corresponding to tangling and missing classification tasks. After each epoch, model performance was evaluated on the test set, predictions were stored with ground-truth labels, and key metrics were recorded. For reproducibility and the ability to resume training, checkpoints of the shared MLP and the classification heads were saved after each epoch, while the frozen backbone was restored directly from pretrained weights when needed.

%% file: methodology-llm-based-analysis-experimental-setup.tex
\subsubsection{Experimental Setup}
\label{subsubsec:llm-based-analysis-experimental-setup}

\begin{figure*}[h]
  \centering
  \includegraphics[width=\textwidth]{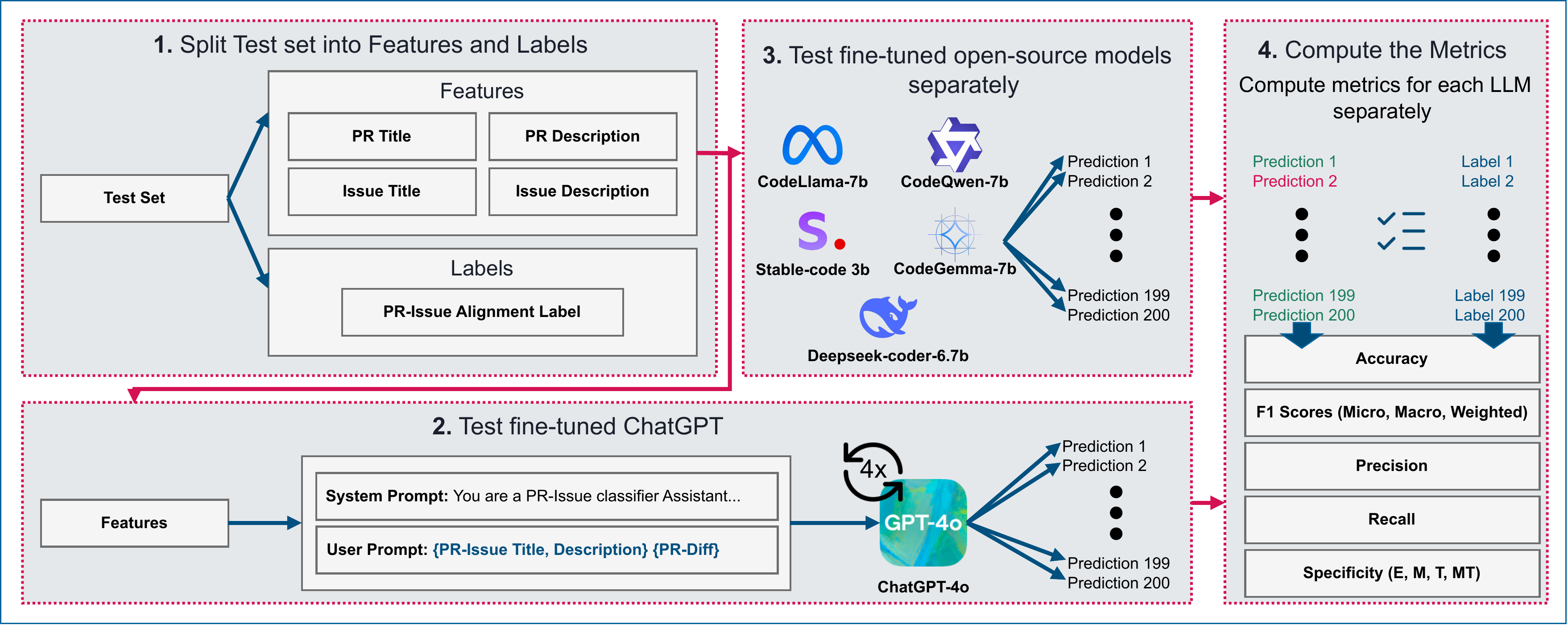}
\caption{LLM-Based Analysis Experimental Setup}
\label{fig:llm-based-analysis-experimental-setup}
\end{figure*}

Following the fine-tuning process, we evaluated both GPT-4o and the open-source models within a controlled experimental pipeline. The test set was divided into a feature set comprising the PR title, PR description, issue title, issue description, and the PR–issue alignment label, as illustrated in Step~1 of Fig.~\ref{fig:llm-based-analysis-experimental-setup}. These features were then mapped into system and user prompts, as given in Table~\ref{tab:pr_issue_alignment_prompt}, to query fine-tuned GPT-4o. To maximize output consistency and minimize stochastic variability, the temperature parameter was fixed at zero, and each data pair was submitted four times during experimentation, with the most frequent label among the outputs considered as the final predicted label. While this setup substantially reduced randomness in model behavior, complete determinism could not be guaranteed due to inherent hardware and batch-level nondeterminism. The corresponding labels were then recorded across epochs, as presented in Step~2 of Fig.~\ref{fig:llm-based-analysis-experimental-setup}.


Subsequently, we evaluated the open-source models CodeLLaMA-7B, CodeQwen1.5-7B, StableCode-3B, CodeGemma-7B, and Deepseek-Coder-6.7B separately. Predictions were recorded for each model across epochs, as shown in Step~3 of Fig.~\ref{fig:llm-based-analysis-experimental-setup}. Since the language modeling head—including the linear and softmax layers that define sampling temperature—was removed during architectural modification (Section~\ref{subsubsec:fine-tuning-open-source-models}), these models operated deterministically. Consequently, there was no temperature parameter or stochastic decoding involved, and each test sample was evaluated once, as repeated querying was unnecessary to estimate statistical confidence.

After collecting the outputs across epochs for all models, we selected the epochs demonstrating the highest accuracy on test set for further evaluation. On these selected epochs, we computed the final evaluation metrics, including accuracy, F1-scores (micro, macro, and weighted) precision, recall, and specificity (Exact, Missing, Tangling, Missing and Tangling).

%% file: methodology-interpretability-analysis.tex
\subsection{Interpretability Analysis}
\label{subsec:interpretability-analysis}
%
To investigate the influence of different PR–issue fields on classification outcomes, we conducted a SHAP \citep{lundberg2017unified} analysis on CodeLlama-7B, the open-source LLM that demonstrated the highest classification accuracy. SHAP was chosen due to its model-agnostic nature, ability to provide consistent and theoretically sound attributions, and its suitability for multi-head classification tasks. Since the LLM backbone is frozen, SHAP is particularly appropriate as it evaluates feature contributions through the trainable heads without requiring gradient flow through the frozen layers. Unlike LIME or IG, SHAP allows for a unified framework that can handle interactions between input features and output predictions, making it suitable for our fine-tuned LLMs.

Since the fine-tuned LLMs includes two binary classification heads, \textit{isMissing} and \textit{isTangling}. For each head, SHAP values were computed independently for each filed, and these values were visualized using Beeswarm plots, which illustrate the distribution and importance of feature contributions across instances. The mean of the absolute SHAP values (Mean(\textbar SHAP\textbar)) was then calculated to quantify the overall impact of each field on the model's decisions. This approach offers a comprehensive interpretability assessment and highlights which PR–issue fields most strongly drive the model's predictions.


%% file: results.tex
\section{Results}
\label{sec:results}
In this Section, we present the results obtained from both LLM-based and interpretability analysis. This Section presents the results obtained from both LLM-based and interpretability analyses. Section \ref{subsec:llm-based-analysis-results} reports the outcomes of the LLM-based classification analysis and provides answers to RQ1 and RQ2. Subsequently, Section \ref{subsec:interpretability-analysis-results} presents the findings of the SHAP analysis results and answers RQ3.

\input{results-llm-based-analysis}
\input{results-interpretability-analysis}

%% file: results-llm-based-analysis.tex
\subsection{LLM-Based Analysis Results}
\label{subsec:llm-based-analysis-results}
This Section presents the results of LLM-based analysis and compares the performance of fine-tuned models with the findings of \citet{bilsen} on the PR-issue alignment classification task. We first address RQ1 by comparing the performance of our fine-tuned models against the prompting approach from previous studies, evaluating metrics including accuracy, F1-score (micro, macro, and weighted), precision, recall, and specificity (Exact, Missing, and Tangling) as detailed in Section \ref{subsubsec:assessment-of-performance-metrics}. Subsequently, to address RQ2, we evaluate the performance of each of the fine-tuned LLMs (CodeLlama-7B, CodeQwen1.5-7B, StableCode-3B, CodeGemma-7B, Deepseek-Coder-6.7B, and GPT-4o) to identify which model is most effective for this specific task, as further outlined in Section \ref{subsubsec:performance-evaluation-of-fine-tuned-llms}.

\begin{figure}[p]
\centering
    \begin{subfigure}[t]{0.92\textwidth}
        \centering
        \includegraphics[width=\linewidth,height=0.18\textheight,keepaspectratio]{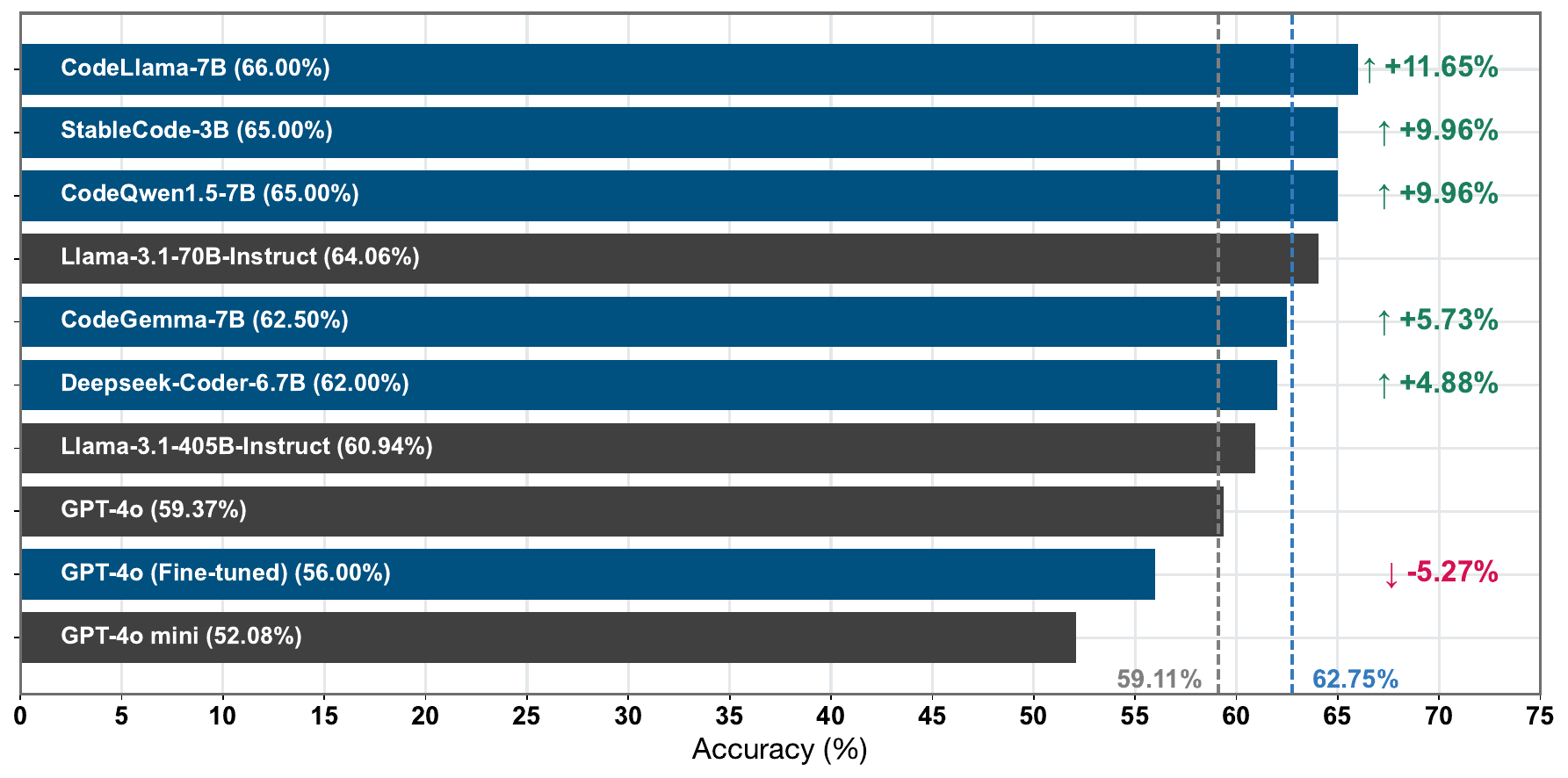}
        \caption{Accuracy of fine-tuned and base LLMs}
        \label{fig:accuracy}
    \end{subfigure}
    \par\vspace{0.5em}
    \begin{subfigure}[t]{0.92\textwidth}
        \centering
        \includegraphics[width=\linewidth,height=0.18\textheight,keepaspectratio]{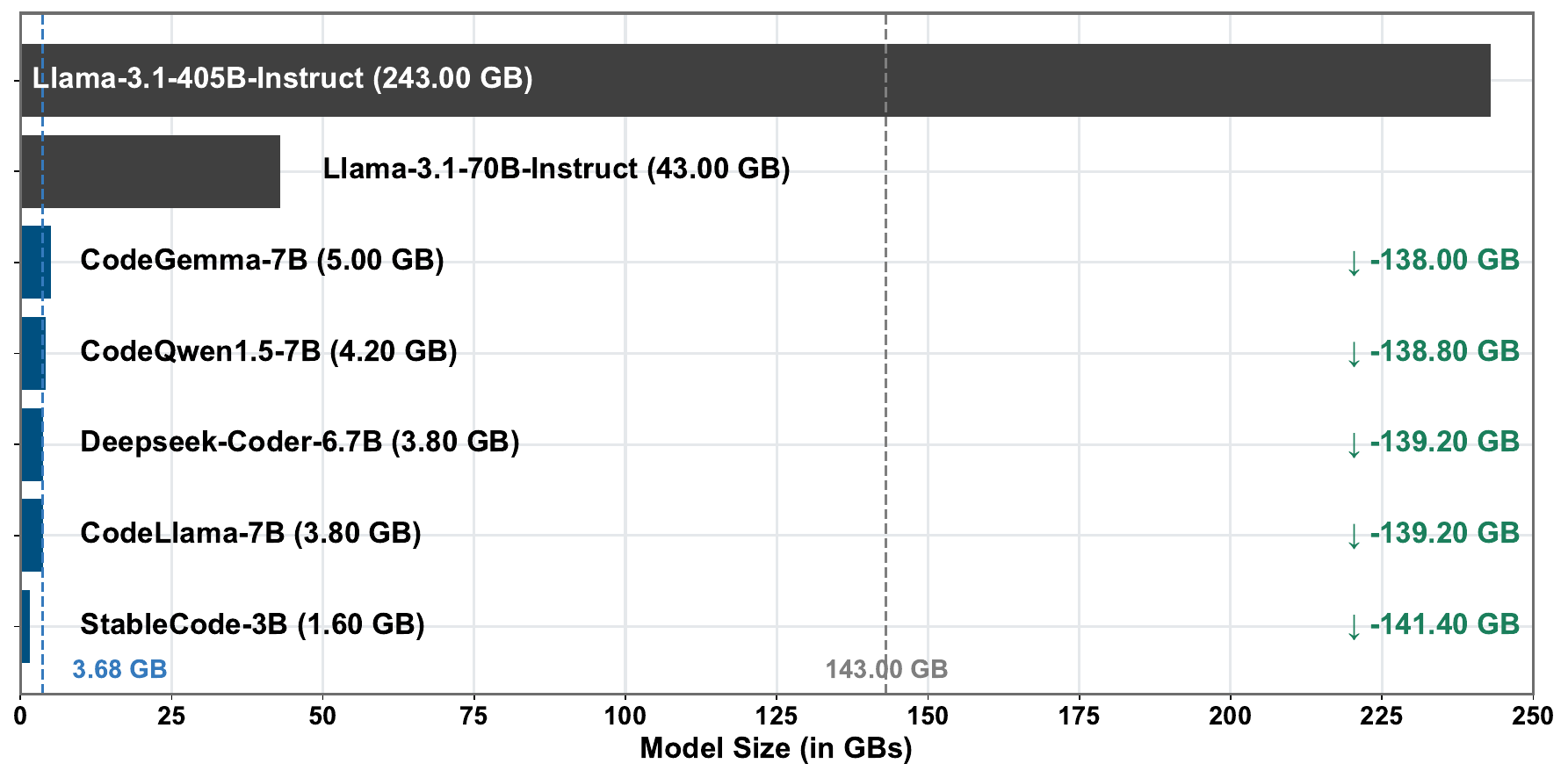}
        \caption{Model size (GB) of fine-tuned and base LLMs}
        \label{fig:model_size}
    \end{subfigure}
    \par\vspace{0.5em}
    \begin{subfigure}[t]{0.92\textwidth}
        \centering
        \includegraphics[width=\linewidth,height=0.18\textheight,keepaspectratio]{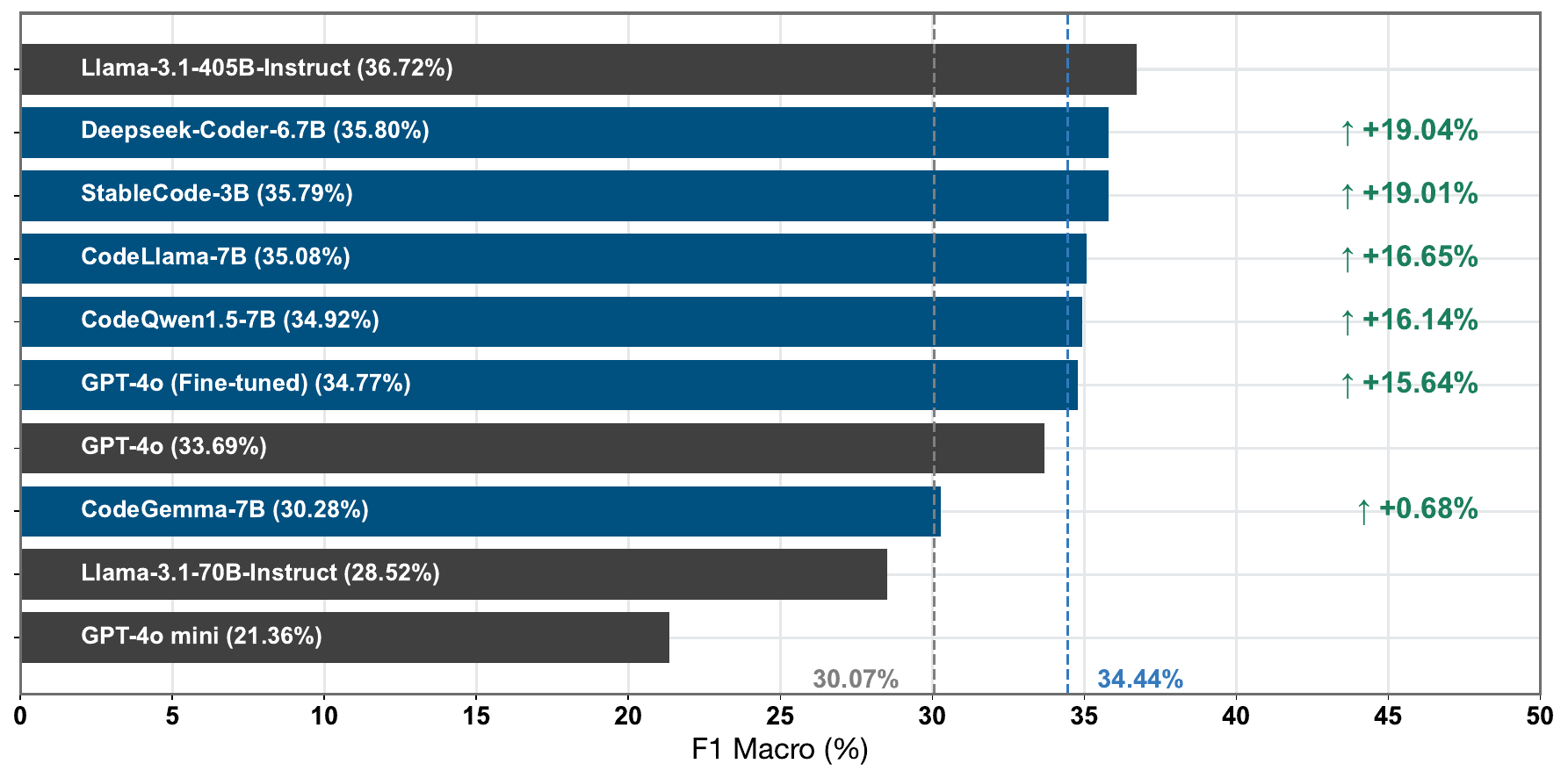}
        \caption{F1-macro Scores of fine-tuned and base LLMs}
        \label{fig:f1_macro}
    \end{subfigure}
    \par\vspace{0.5em}
    \begin{subfigure}[t]{0.46\textwidth}
        \centering
        \includegraphics[width=\linewidth,height=0.16\textheight,keepaspectratio]{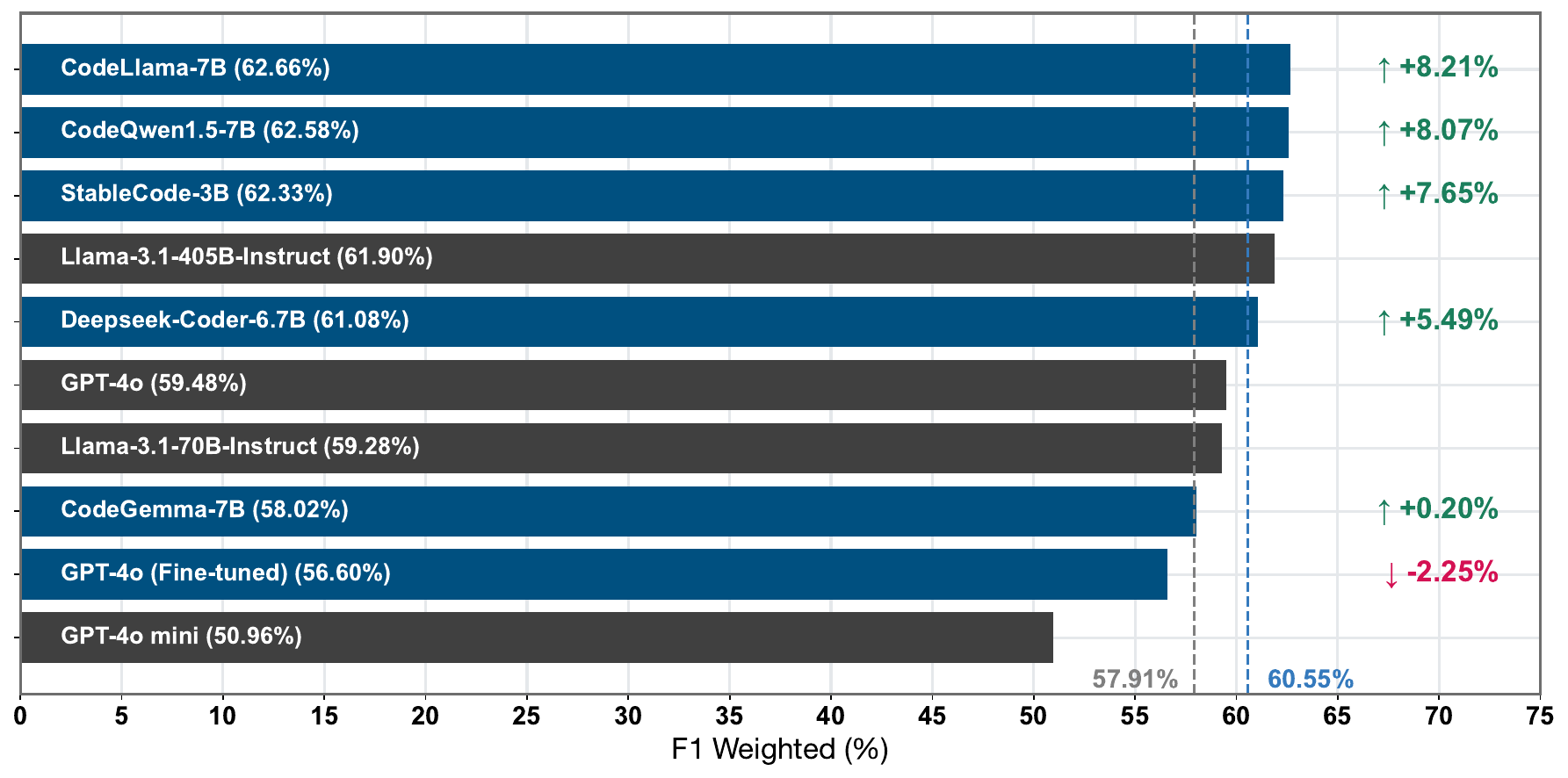}
        \caption{F1-weighted Scores of fine-tuned and base LLMs}
        \label{fig:f1_weighted}
    \end{subfigure}
    \hfill
    \begin{subfigure}[t]{0.46\textwidth}
        \centering
        \includegraphics[width=\linewidth,height=0.16\textheight,keepaspectratio]{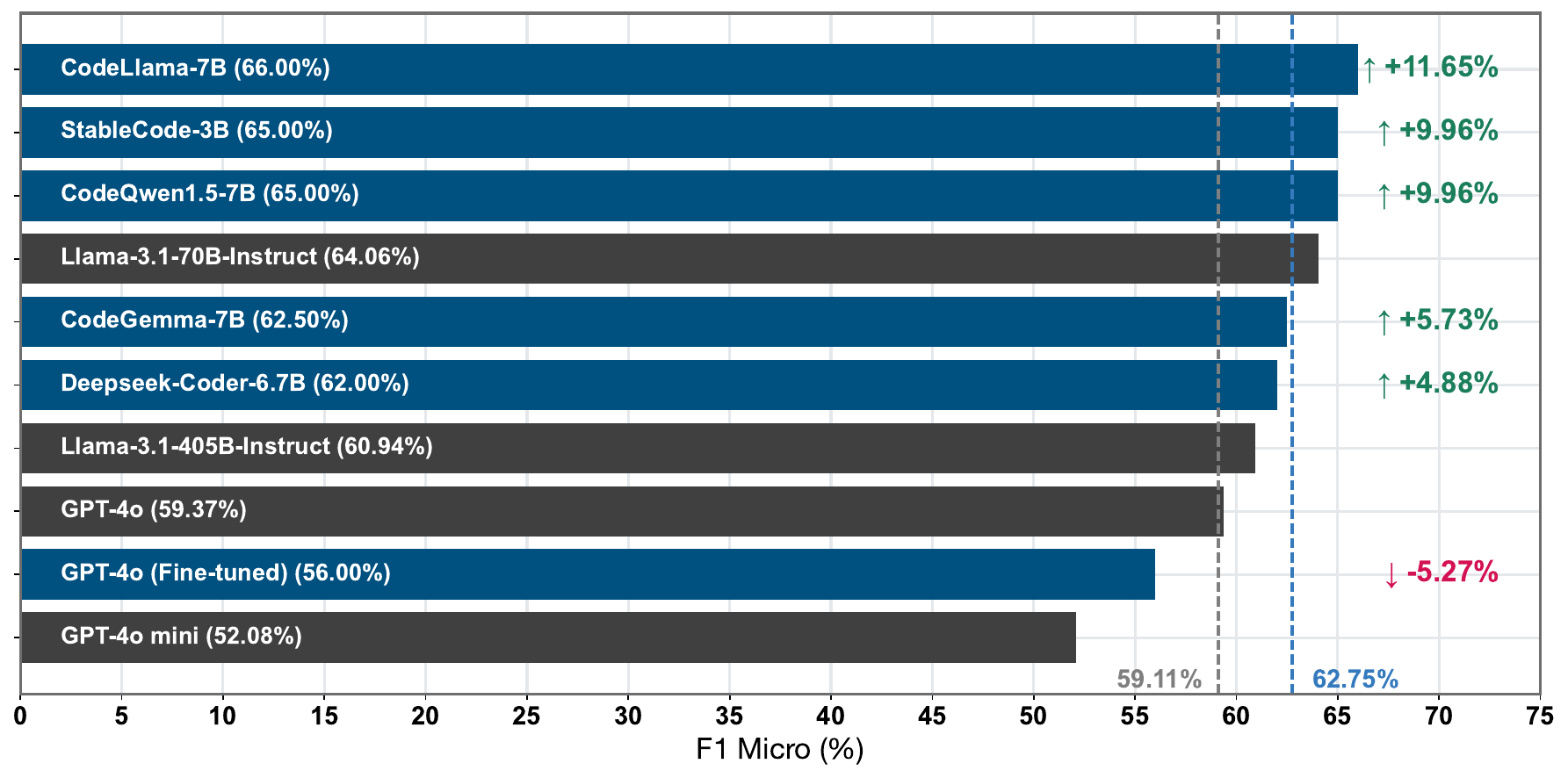}
        \caption{F1-micro Scores of fine-tuned and base LLMs}
        \label{fig:f1_micro}
    \end{subfigure}
    \par\vspace{0.6em}
    \includegraphics[width=0.42\textwidth]{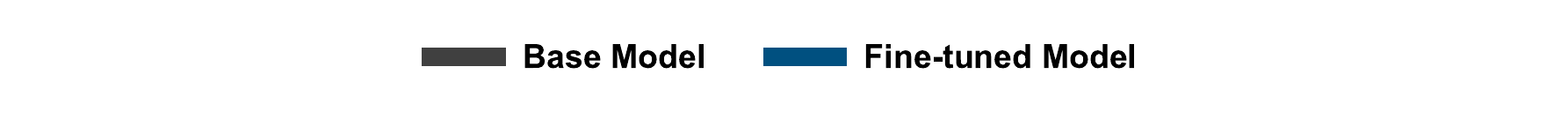}
    \caption{Comparison of fine-tuned and base LLMs in terms of Accuracy, F1-macro, F1-micro, F1-weighted, and model size (GB).}
    \label{fig:metrics_1}
\end{figure}

\begin{figure}[p]
\centering
    \begin{subfigure}[t]{0.92\textwidth}
        \centering
        \includegraphics[width=\linewidth,height=0.17\textheight,keepaspectratio]{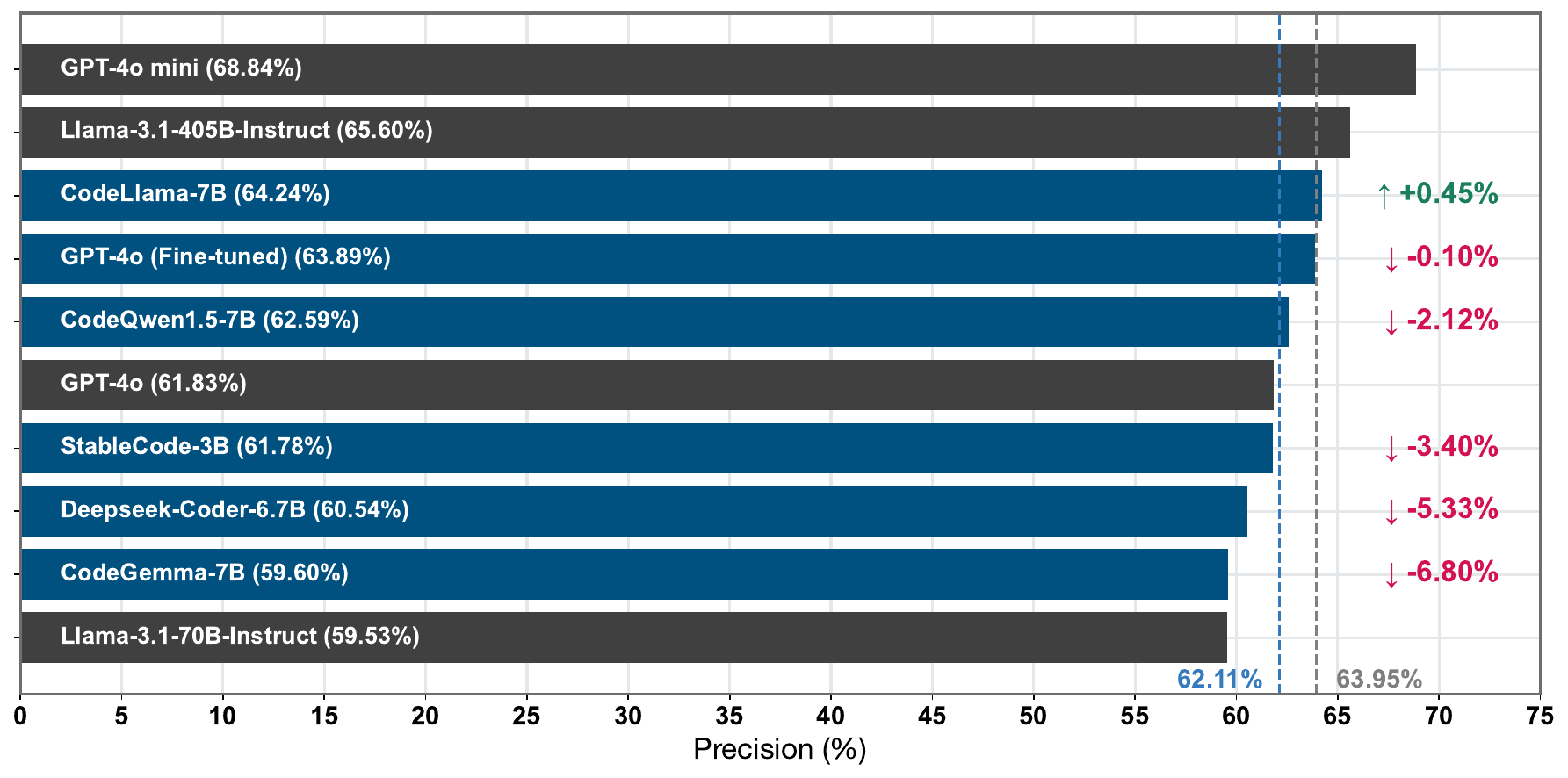}
        \caption{Precision of fine-tuned and base LLMs}
        \label{fig:precision}
    \end{subfigure}
    \par\vspace{0.5em}
    \begin{subfigure}[t]{0.92\textwidth}
        \centering
        \includegraphics[width=\linewidth,height=0.17\textheight,keepaspectratio]{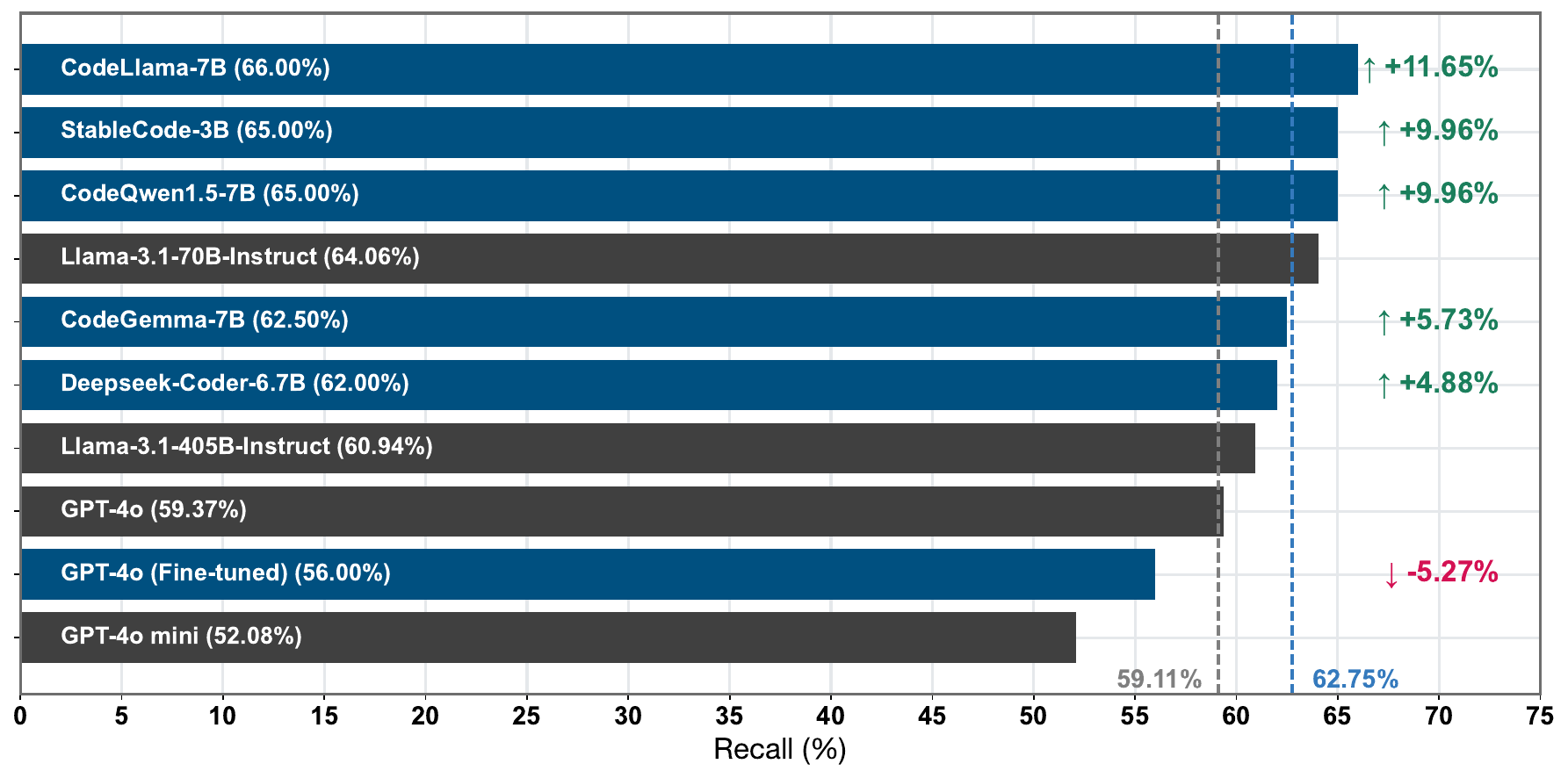}
        \caption{Recall of fine-tuned and base LLMs}
        \label{fig:recall}
    \end{subfigure}
    \par\vspace{0.5em}
    \begin{subfigure}[t]{0.46\textwidth}
        \centering
        \includegraphics[width=\linewidth,height=0.15\textheight,keepaspectratio]{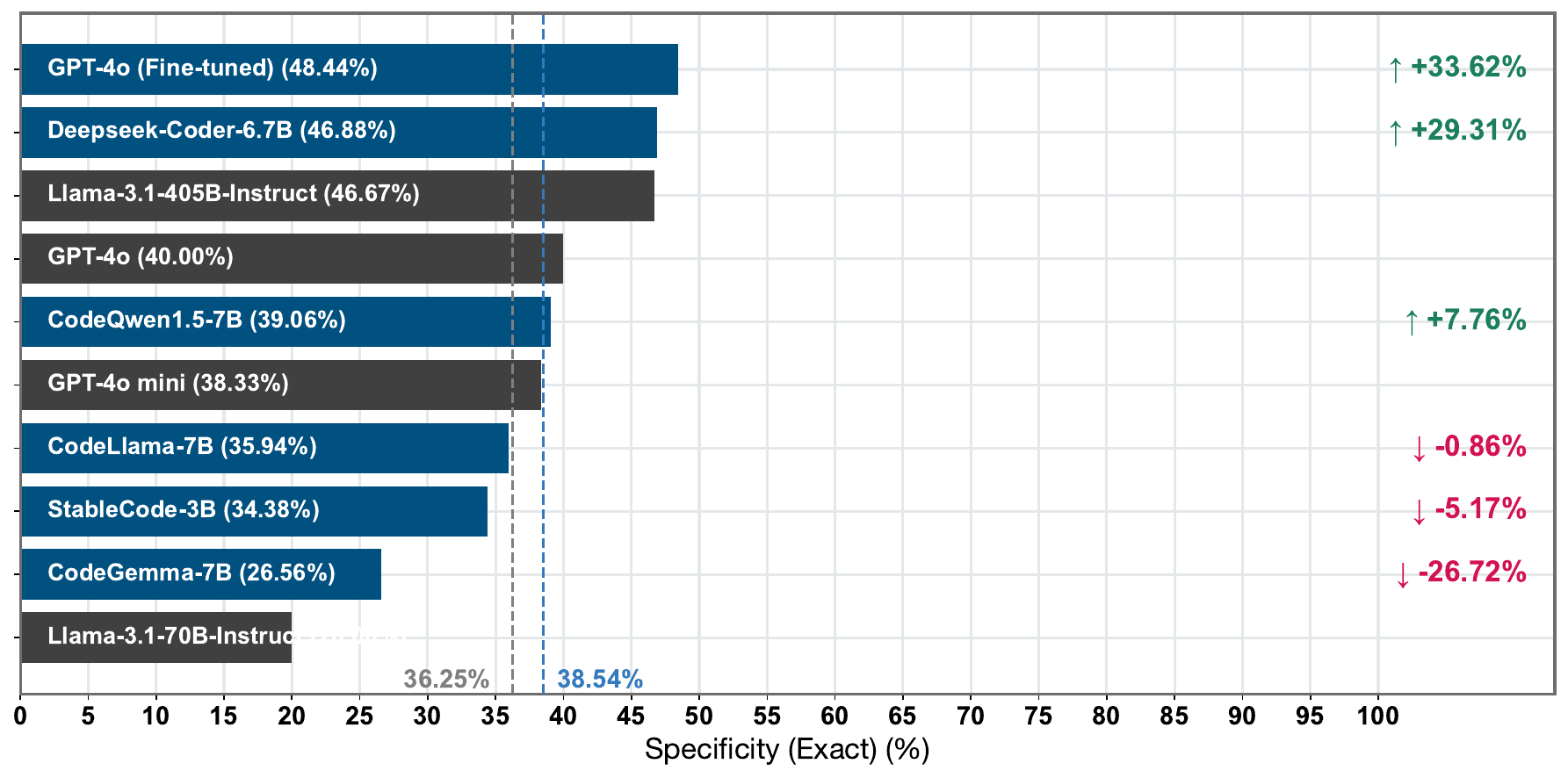}
        \caption{Specificity (Exact) of fine-tuned and base LLMs}
        \label{fig:specificity_E}
    \end{subfigure}
    \hfill
    \begin{subfigure}[t]{0.46\textwidth}
        \centering
        \includegraphics[width=\linewidth,height=0.15\textheight,keepaspectratio]{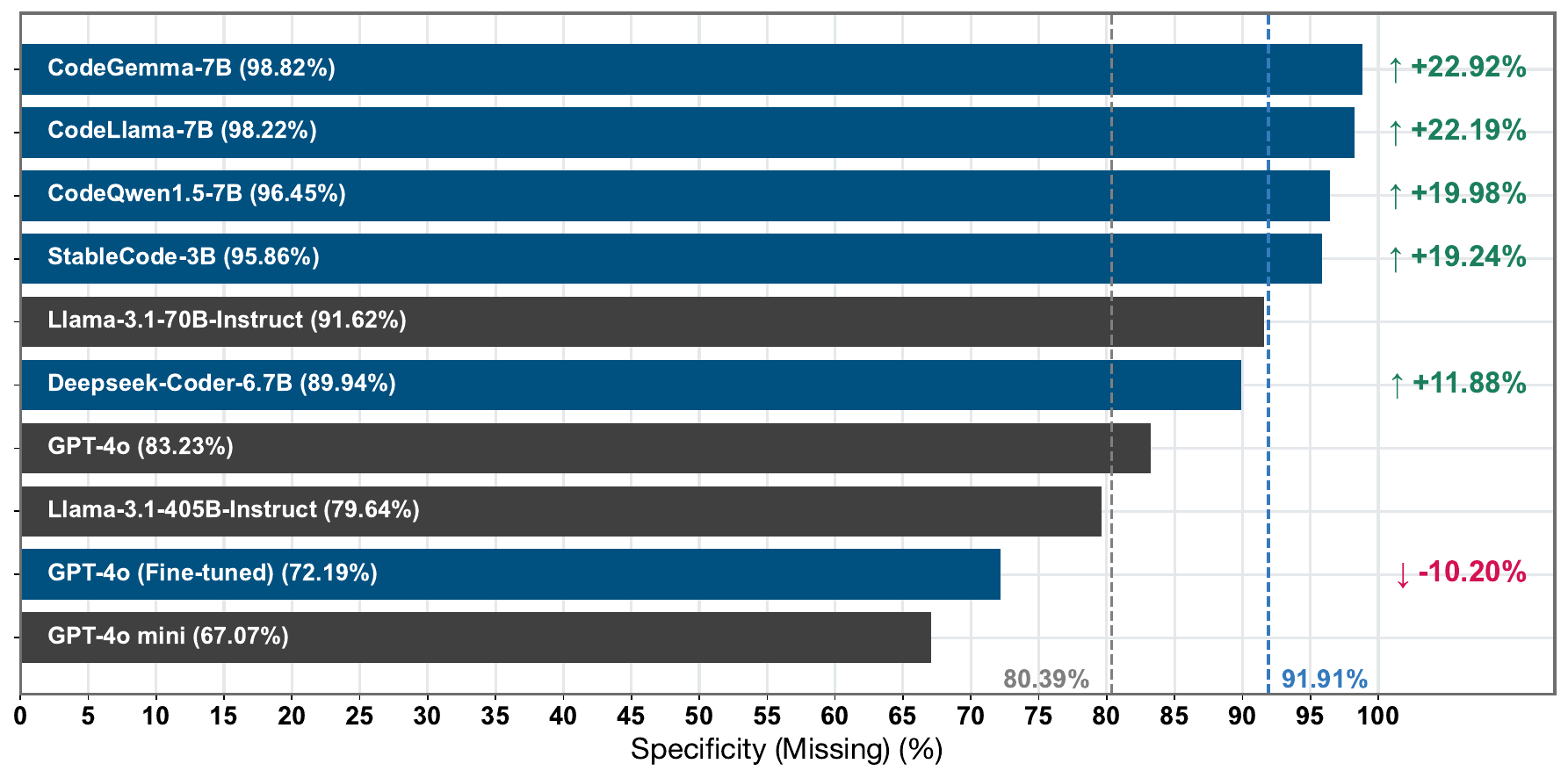}
        \caption{Specificity (Missing) of fine-tuned and base LLMs}
        \label{fig:specificity_M}
    \end{subfigure}
    \par\vspace{0.5em}
    \begin{subfigure}[t]{0.46\textwidth}
        \centering
        \includegraphics[width=\linewidth,height=0.15\textheight,keepaspectratio]{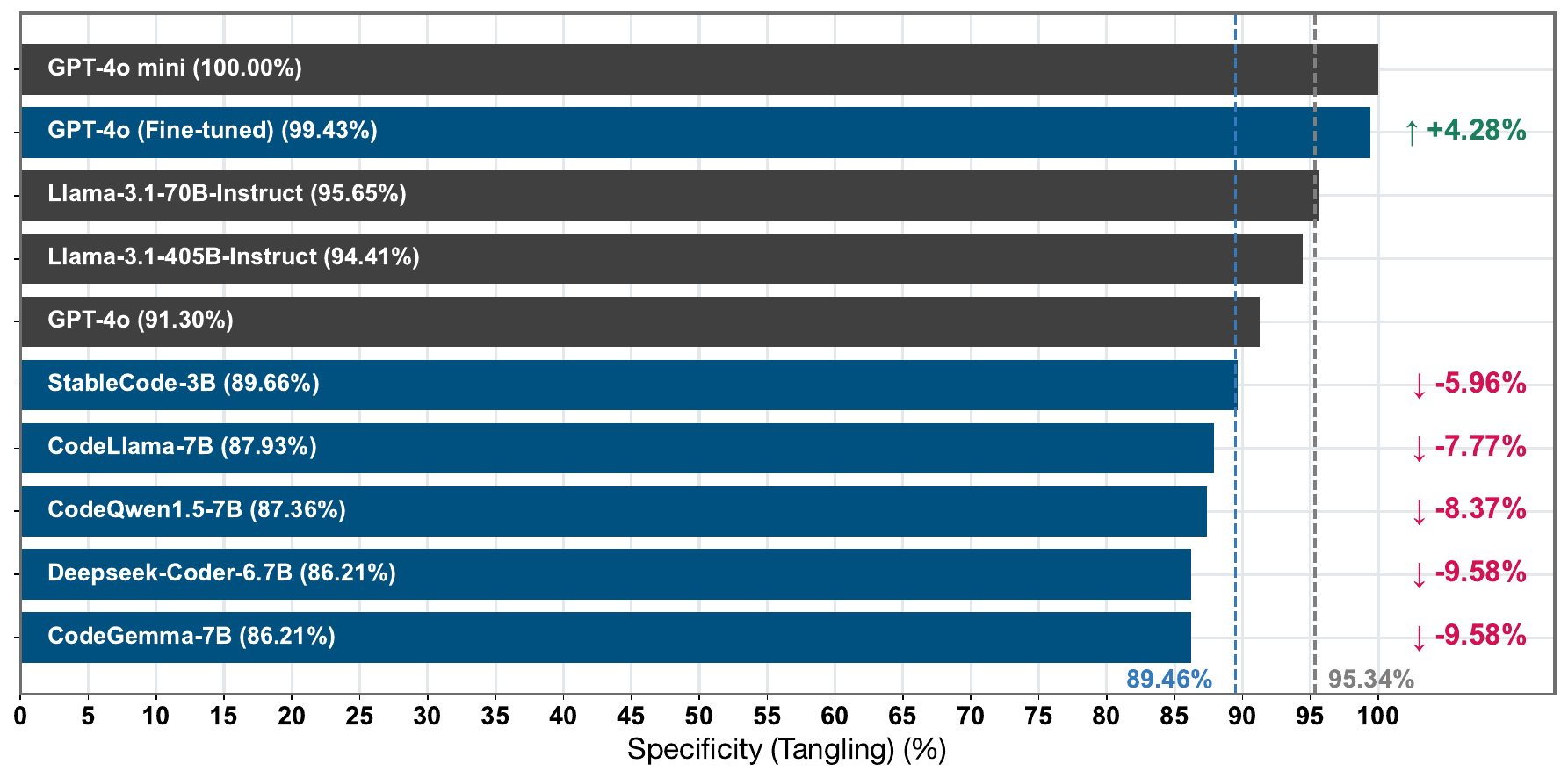}
        \caption{Specificity (Tangling) of fine-tuned and base LLMs}
        \label{fig:specificity_T}
    \end{subfigure}
    \hfill
    \begin{subfigure}[t]{0.46\textwidth}
        \centering
        \includegraphics[width=\linewidth,height=0.15\textheight,keepaspectratio]{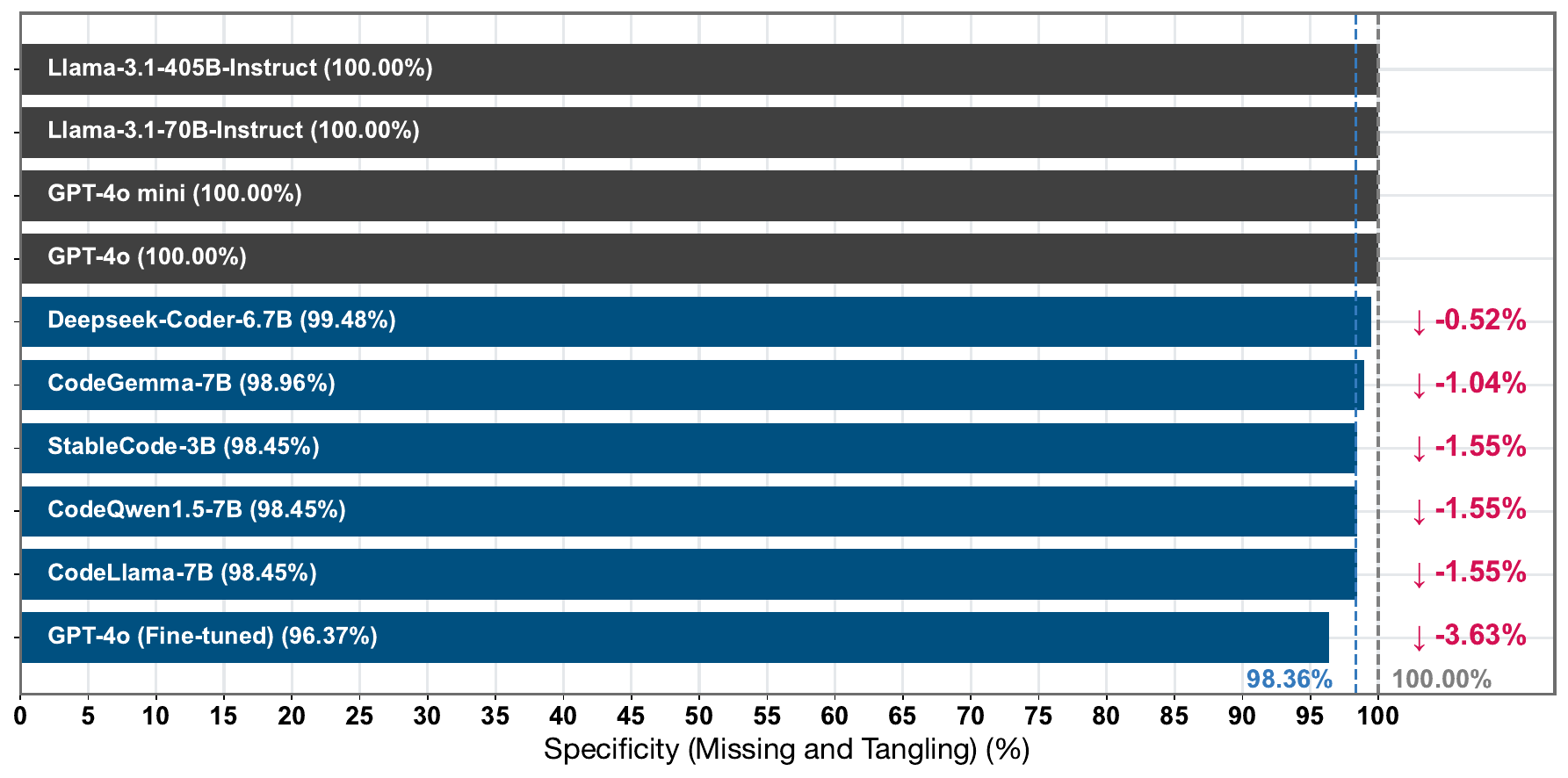}
        \caption{Specificity (Missing and Tangling) of fine-tuned and base LLMs}
        \label{fig:specificity_MT}
    \end{subfigure}
    \par\vspace{0.6em}
    \includegraphics[width=0.42\textwidth]{v1_model_legend.pdf}
    \caption{Comparison of fine-tuned and base LLMs across Precision, Recall, and various Specificity metrics.}
    \label{fig:metrics_2}
\end{figure}

Fig. \ref{fig:metrics_1}  and  Fig. \ref{fig:metrics_2} demonstrates the evaluation metrics include accuracy, size, F1-score (micro, macro, and weighted), precision, recall, and specificity. In these plots, the LLMs fine-tuned in this study are compared against the base models reported in previous work, with gray bars indicating prior performance and blue bars representing the results of our fine-tuned models. Each plot includes vertical dashed lines that denote the average performance of both the fine-tuned models and the base models, distinguished by their respective colors. Fine-tuned models explicitly highlight the difference from the average performance of the base models, thereby illustrating the relative improvement achieved through fine-tuning across the examined metrics.

\subsubsection{Performance Comparison with Previous Approach}
\label{subsubsec:assessment-of-performance-metrics}
The performance of the fine-tuned LLMs and base models from previous work \citep{bilsen} is evaluated across specified metrics, focusing on both average values and the outcomes of the best-performing models. The comparison of averages provides a clear indication of overall improvement achieved through fine-tuning, while the analysis of individual model results highlights which fine-tuned LLMs demonstrate the greatest advancement relative to average of base LLMs. The reported standard deviation is calculated only across the fine-tuned models to show variability among them.

\begin{enumerate}[label=\textbf{\arabic*.}]
   \item \textbf{Accuracy:} Fine-tuned models achieved an average accuracy of 62.75\%, a 6.15\% improvement over the base models' average of 59.11\%. CodeLlama-7B attained the highest accuracy at 66.00\%, showing a 3.02\% gain over the strongest base model, Llama-3.1-70B-Instruct (64.06\%). Other top fine-tuned performers were CodeQwen1.5-7B and StableCode-3B with 65.00\% accuracy. The standard deviation of 3.66\% indicated moderate variability in performance, while GPT-4o showed a slight decrease (56.00\%).

   \item \textbf{Size:} Model storage size comparison excludes proprietary models (GPT-4o, GPT-4o). Fine-tuned models achieved an impressive average size of 3.68 GB, a 97.42\% reduction compared to the base models' average of 143 GB. The most compact fine-tuned model, StableCode-3B (1.6 GB), achieved a 96.3\% decrease in storage requirements compared to the smallest base model, Llama-3.1-70B-Instruct (43 GB). This drastic reduction emphasizes the practicality of fine-tuned models for deployment.

   \item \textbf{F1-Macro:} Fine-tuned models showed performance gains in F1-macro (average 34.44\%), representing a 14.53\% improvement over base models (30.07\%). Deepseek-Coder-6.7B attained the highest F1-macro score (35.80\%), followed closely by StableCode-3B (35.79\%). While the strongest base model, Llama-3.1-405B-Instruct (36.72\%), marginally outperformed the best fine-tuned model, the consistent gains across architectures (Standard Deviation: 1.96\%) underscore the effectiveness of fine-tuning for generalization across imbalanced classes.

   \item \textbf{F1-Weighted:} Fine-tuned models generally surpassed base models in weighted F1 (average 60.55\% vs. 57.90\%), achieving an average improvement of 4.57\%. CodeLlama-7B recorded the highest weighted F1 at 62.66\%, showing a 0.76\% gain over the strongest base model, Llama-3.1-405B-Instruct (61.90\%). Stable improvements were noted across the fine-tuned models (Standard Deviation: 2.61\%).

   \item \textbf{F1-Micro:} Consistent with Accuracy, fine-tuned models demonstrated a 6.15\% improvement in F1-micro (average 62.75\% vs. 59.11\% for base models). CodeLlama-7B led with 66.00\%, achieving a 1.94\% gain over the top base model, Llama-3.1-70B-Instruct (64.06\%), highlighting the positive impact of fine-tuning on overall classification performance.

   \item \textbf{Precision:} Fine-tuned models showed a slight average decrease in precision (62.11\% vs. 63.95\% for base models), indicating a tradeoff with recall. CodeLlama-7B achieved the highest precision among fine-tuned models at 64.24\%. However, the best base model, GPT-4o mini (68.84\%), maintained a superior precision score, emphasizing the precision/recall balance.

   \item \textbf{Recall:} Fine-tuned models exhibited higher recall (average 62.75\% vs. 59.11\%), a 6.17\% improvement. CodeLlama-7B again led with 66.00\% recall, securing a 1.94\% gain over the strongest base model, Llama-3.1-70B-Instruct (64.06\%), confirming the efficacy of fine-tuning in enhancing the capture of positive instances.

   \item \textbf{Specificity (Exact):} Fine-tuned models achieved an average specificity (Exact) of 38.54\%, a 6.33\% improvement. GPT-4o, among the fine-tuned models, achieved the highest specificity at 48.44\%, surpassing the strongest base model, Llama-3.1-405B-Instruct (46.67\%), by 1.77\% in detecting exact PR-issue alignment.

   \item \textbf{Specificity (Missing):} Fine-tuned models demonstrated a significant average improvement of 14.33\% (average 91.91\%) in detecting missing elements. CodeGemma-7B achieved the highest score at 98.82\%, which is a 7.85\% gain over the best base model, Llama-3.1-70B-Instruct (91.62\%).

   \item \textbf{Specificity (Tangling):} Fine-tuned models showed an average decrease of 6.16\% (average 89.47\%) compared to base models (95.34\%). GPT-4o, however, maintained competitive performance at 99.43\%, only 0.57\% below the perfect performance of the strongest base model, GPT-4o mini (100.00\%).

   \item \textbf{Specificity (Missing and Tangling):} Fine-tuned models averaged 98.36\%, a minor reduction of 1.64\% compared to the base models' perfect 100\% average. Deepseek-Coder-6.7B was the top fine-tuned model at 99.48\%, demonstrating that fine-tuned models maintain high robustness in combined misalignment detection.
\end{enumerate}

\textbf{Overall Evaluation:} The overall findings indicate that fine-tuned models perform better than base models across most evaluation metrics. On average, fine-tuning resulted in improvements of 6.15\% in accuracy, 6.15\% in F1-micro, 14.69\% in F1-macro, 4.56\% in F1-weighted, 6.15\% in recall, 6.33\% in specificity (Exact), and 14.33\% in specificity (Missing), while also occupying 97.43\% less size compared to base models. These consistent gains demonstrate the advantage of fine-tuning for enhancing PR–issue alignment classification. On the other hand, base models exhibited better performance in certain areas, 2.97\% in precision, 1.67\% in specificity (Missing and Tangling), and 6.56\% in specificity (Tangling). This contrast highlights that while fine-tuned models generally deliver superior results, base models retain some advantages in precision-oriented and tangling-related specificity tasks.

When comparing the best-performing fine-tuned models with the best base models for each metric, fine-tuned models generally show superior performance in most evaluation areas, expressed as percentage improvements or declines. For accuracy, CodeLlama-7B outperforms Llama-3.1-70B-Instruct by 3.02\%, while F1-micro also improves by 3.02\%. In F1-macro, Deepseek-Coder-6.7B slightly underperforms Llama-3.1-405B by 1.64\%, whereas F1-weighted sees a 0.76\% gain. Precision declines by 4.6\% for the best fine-tuned model compared to best performing base model (Llama-3.1-70B Instruct), and recall improves by 1.94\%. Specificity (Exact) increases by 1.77\%, Specificity (Missing) rises by 7.2\%, Specificity (Missing and Tangling) drops by 0.52\%, and Specificity (Tangling) decreases by 0.57\%. In terms of storage space, StableCode-3B occupies far less memory than the smallest base model (Llama-3.1-70B), showing a 41.4 GB reduction. Overall, comparing the best fine-tuned and best base models for each metric highlights that fine-tuning generally enhances PR-issue alignment performance, although certain metrics, especially precision and some specificity measures, exhibit minor declines.

\tcolorbox[colback=figure-soft-white,    
           colframe=figure-navy-blue] 

\textbf{RQ1: How do fine-tuned LLMs perform in classifying PR–issue alignment compared to previous approaches?}\\
Fine-tuned LLMs consistently demonstrate improved performance in classifying PR-issue alignment when compared to base models in previous approach. The overall evaluation indicates that fine-tuning yields notable gains across most evaluation metrics, with average improvements of 6.15\% in accuracy and F1-micro, 14.69\% in F1-macro, and 6.15\% in recall, alongside over 95\% reductions in model size. When focusing on the best-performing models for each metric, fine-tuned versions maintain superiority in accuracy, recall, and most specificity measures, while exhibiting modest declines in precision and certain tangling-related specificity tasks. These findings suggest that fine-tuned LLMs not only enhance classification robustness but also achieve greater efficiency in terms of resource utilization. Collectively, the results establish fine-tuning as an effective strategy for advancing PR--issue alignment classification, while also highlighting areas where base models may retain specific advantages.
\endtcolorbox

\subsubsection{Performance Evaluation of Fine-Tuned LLMs}
\label{subsubsec:performance-evaluation-of-fine-tuned-llms}

To determine the best performing fine-tuned LLM for the PR-issue alignment task, each model's performance and rank is evaluated based on accuracy, size , F1-score (micro, macro, and weighted), precision, recall, specificity and storage size. The performance of each fine-tuned LLM is evaluated and compared against the other fine-tuned models, with Table \ref{table:comparison-table} providing a detailed breakdown of each model’s performance metrics and their corresponding relative ranks. This comparative analysis enables the identification of the top-performing LLMs across different metrics. Furthermore, this section proposes a consolidated  overall scoring system to identify the single best-ranking LLM based on its cumulative performance across all measures.

\begin{table}[htbp]
\centering
\scriptsize
\setlength{\tabcolsep}{4pt} 
\begin{tabular}{l|c|c|c|c|c|c}
\hline
Metric & GPT-4o & CodeLlama-7B & CodeQwen1.5-7B & StableCode-3B & CodeGemma-7B & Deepseek-Coder-6.7B \\
\hline
Accuracy (\%) & 56.00\textsuperscript{6th} & \textbf{66.00\textsuperscript{1st}} & 65.00\textsuperscript{2nd} & 65.00\textsuperscript{2nd} & 62.50\textsuperscript{4th} & 62.00\textsuperscript{5th} \\
F1-micro    (\%) & 56.00\textsuperscript{6th} & \textbf{66.00\textsuperscript{1st}} & 65.00\textsuperscript{2nd} & 65.00\textsuperscript{2nd} & 62.50\textsuperscript{4th} & 62.00\textsuperscript{5th} \\
F1-macro (\%) & 34.77\textsuperscript{5th} & 35.08\textsuperscript{3rd} & 34.92\textsuperscript{4th} & 35.79\textsuperscript{2nd} & 30.58\textsuperscript{6th} & \textbf{35.80\textsuperscript{1st}} \\
F1-weighted (\%) & 56.60\textsuperscript{6th} & \textbf{62.66\textsuperscript{1st}} & 62.58\textsuperscript{2nd} & 62.33\textsuperscript{3rd} & 58.02\textsuperscript{5th} & 61.08\textsuperscript{4th} \\
Precision (\%) & 63.89\textsuperscript{2nd} & \textbf{64.24\textsuperscript{1st}} & 62.59\textsuperscript{3rd} & 61.78\textsuperscript{4th} & 59.60\textsuperscript{6th} & 60.54\textsuperscript{5th} \\
Recall (\%) & 56.00\textsuperscript{6th} & \textbf{66.00\textsuperscript{1st}} & 65.00\textsuperscript{2nd} & 65.00\textsuperscript{2nd} & 62.50\textsuperscript{4th} & 62.00\textsuperscript{5th} \\
Spec-Exact (\%) & \textbf{48.44\textsuperscript{1st}} & 35.94\textsuperscript{4th} & 39.06\textsuperscript{3rd} & 34.38\textsuperscript{5th} & 26.56\textsuperscript{6th} & 46.88\textsuperscript{2nd} \\
Spec-Missing (\%) & 72.19\textsuperscript{6th} & 98.22\textsuperscript{2nd} & 96.45\textsuperscript{3rd} & 95.86\textsuperscript{4th} & \textbf{98.82\textsuperscript{1st}} & 89.94\textsuperscript{5th} \\
Spec-MT (\%) & 96.37\textsuperscript{6th} & 98.45\textsuperscript{3rd} & 98.45\textsuperscript{3rd} & 98.45\textsuperscript{3rd} & 98.96\textsuperscript{2nd} & \textbf{99.48\textsuperscript{1st}} \\
Spec-Tangling (\%) & \textbf{99.43\textsuperscript{1st}} & 87.93\textsuperscript{3rd} & 87.36\textsuperscript{4th} & 89.66\textsuperscript{2nd} & 86.21\textsuperscript{5th} & 86.21\textsuperscript{5th} \\
Size (GB) & - & 3.8\textsuperscript{2nd} & 4.2\textsuperscript{4th} & \textbf{1.6\textsuperscript{1st}} & 5.0\textsuperscript{5th} & 3.8\textsuperscript{2nd} \\
\hline
\end{tabular}
\caption{Fine-Tuned LLMs classification performance with ranks}
\label{table:comparison-table}
\end{table}

\begin{enumerate}[label=\textbf{\arabic*.}]
   \item \textbf{GPT-4o:} This model, GPT-4o, demonstrates suboptimal performance, which is notable given the high financial and computational costs associated with fine-tuning a proprietary model. In terms of overall classification performance, the model consistently ranked sixth in key metrics, including Accuracy (56.00\%), F1-micro (56.00\%), F1-weighted (56.60\%), and Recall (56.00\%). While it did secure a second-place rank for Precision (63.89\%) and a first-place rank for both Specificity (Exact) (48.44\%) and Specificity (Tangling) (99.43\%), these strengths are counterbalanced by a sixth-place rank in Specificity (Missing) (72.19\%) and Specificity (Missing and Tangling) (96.37\%). The model's underperformance in these critical metrics indicates that its fine-tuning did not yield superior results compared to smaller, open-source LLMs.
    \item \textbf{CodeLlama-7B:} The CodeLlama-7B model demonstrated a strong performance profile, securing a first-place ranking in several key metrics, including Accuracy (66.00\%), F1-micro (66.00\%), F1-weighted (62.66\%), Precision (64.24\%), and Recall (66.00\%). While it ranked third in F1-macro (35.08\%) and Specificity (Tangling) (87.93\%), and fourth in Specificity (Exact) (35.94\%), its balanced performance across these primary classification metrics is noteworthy. This robust capability is particularly significant given its compact size of 3.8GB, which ranks it as the second smallest model evaluated. The combination of its superior performance in core metrics and its highly efficient size positions CodeLlama-7B as an effective solution for resource-constrained environments, outperforming other fine-tuned LLMs.
   \item \textbf{CodeQwen1.5-7B:} CodeQwen1.5-7B demonstrated competitive performance, achieving a second-place ranking in several key metrics, including Accuracy (65.00\%), F1-micro (65.00\%), F1-weighted (62.58\%), and Recall (65.00\%). The model's Precision (62.59\%) and Specificity (Exact) (39.06\%) scores secured a third-place rank, while its Specificity (Missing) and Specificity (Missing and Tangling) metrics also placed it third with values of 96.45\% and 98.45\%, respectively. It ranked fourth for F1-macro (34.92\%) and Specificity (Tangling) (87.36\%). With a size of 4.2GB, ranking fourth overall, its performance is notable given its parameter count, particularly when compared to its similar-sized competitor, Deepseek-Coder-6.7B. These results indicate that CodeQwen1.5-7B is a robust model, consistently placing in the top tiers across a variety of classification metrics.
    \item \textbf{CodeGemma-7B:} CodeGemma-7B achieved a fourth-place ranking for Accuracy, F1-micro, and Recall, each with a score of 62.50\%. For F1-weighted, it ranked fifth with 58.02\%, while it was ranked last for Precision (59.60\%) and F1-macro (30.58\%). A notable result for this model is its first-place ranking in Specificity (Missing) at 98.82\%, and its second-place rank for Specificity (Missing and Tangling) at 98.96\%. However, it ranked last in Specificity (Exact) at 26.56\% and fifth in Specificity (Tangling) at 86.21\%. With a size of 5.0GB, which ranks it fifth among the models evaluated, its overall performance profile indicates strengths in specific negative-case metrics, but with comparatively lower performance in general classification metrics such as F1-macro and Precision.
   \item \textbf{StableCode-3B:} StableCode-3B exhibited a strong performance profile, particularly when considering its rank as the most compact model at 1.6GB. The model achieved competitive second-place rankings in Accuracy, F1-micro, and Recall, all at 65.00\%. It further demonstrated notable performance with a second-place ranking in F1-macro (35.79\%) and Specificity (Tangling) (89.66\%), and a third-place ranking for F1-weighted (62.33\%) and Specificity (Missing and Tangling) (98.45\%). However, its performance was less robust in other areas, such as Precision, where it ranked fourth at 61.78\%, and Specificity (Exact), where it ranked fifth at 34.38\%. This indicates a trade-off where its efficiency and general performance on several metrics come with a slight reduction in specific aspects of precision and exact match capabilities. This performance is notable as it suggests the model provides a highly efficient solution that remains competitive with models more than twice its size, making it a viable option for environments with limited computational resources.
   \item \textbf{Deepseek-Coder-6.7B:} This model's performance on the evaluated metrics demonstrates a complex profile of strengths and weaknesses. Deepseek-Coder-6.7B, with a size of 3.8~GB (tied for second), achieved a fifth-place rank in overall Accuracy, F1-micro, and Recall, each at 62.00\%. However, it distinguished itself by securing the top rank in both F1-macro (35.80\%) and Specificity (Missing and Tangling) (99.48\%), indicating a robust ability to handle class imbalance and maintain high performance in tasks involving multi-turn interactions. While its Precision was ranked fifth at 60.54\% and F1-weighted fourth at 61.08\%, it demonstrated notable proficiency in Specificity (Exact) with a second-place rank (46.88\%). This suggests that the model provides an average performance that balances the key metrics of accuracy and recall.
\end{enumerate}
To determine the best performing LLM for PR-issue classification, a composite performance metric was developed by averaging the rank of each model across the performance categories. The notation $R_{\text{metric}}$, represent the rank of a model in that specific performance metric. Ranks of for each metric the LLMs is presented in Table \ref{table:comparison-table}. To mitigate potential bias that could arise from the granularity of individual specificity metrics, the four sub-metrics (Exact, Missing, Tangling, and Missing and Tangling) were consolidated into a single average specificity rank, as defined in Equation \ref{eq:specificity_avg}. The calculation for the overall average rank ($R_{\text{overall}}$) for each LLM is defined as shown in Equation \ref{eq:overall_rank}. During calculation GPT-4o's size rank is ignored since it is not an open source model. Then, LLMs are ordered by their ($R_{\text{overall}}$) from lowest to highest to determine the overall ranking, with the highest ranking LLM being selected as the overall best. This approach provides a fair assessment of a model's proficiency, as it averages the performance rankings across diverse metrics and emphasizes a balanced capability in both general classification performance and nuanced handling of specificity cases.

\begin{equation}
\label{eq:overall_rank}
R_{\text{overall}} = \frac{R_{\text{Accuracy}} + R_{\text{Size}} + R_{\text{F1-micro}} + R_{\text{Recall}} + R_{\text{Precision}} + R_{\text{F1-macro}} + R_{\text{F1-weighted}} + R_{\text{Specificity}_{\text{avg}}}}{8}
\end{equation}

\begin{equation}
\label{eq:specificity_avg}
R_{\text{Specificity}_{\text{avg}}} = \frac{R_{\text{Exact}} + R_{\text{Missing}} + R_{\text{Tangling}} + R_{\text{Missing\ and\ Tangling}}}{4}
\end{equation}
Following the computation of the $R_{\text{overall}}$ scores for all fine-tuned LLMs, CodeLlama-7B emerged with the highest average rank, at 1.42, securing the first position. It was followed by StableCode-3B at 2.32, CodeQwen1.5-7B at 2.70, Deepseek-Coder-6.7B at 3.87, and CodeGemma-7B at 4.87. In the last position, GPT-4o recorded an $R_{\text{overall}}$ score of 5.14.

\tcolorbox[colback=figure-soft-white,    
           colframe=figure-navy-blue] 

\textbf{RQ2: Which fine-tuned LLM achieves the best performance in classifying PR–issue alignment?}

In a detailed analysis of the performance metrics, CodeLlama-7B consistently demonstrated superior performance, ranking first in general metrics like accuracy, F1-micro, and recall. Deepseek-Coder-6.7B excelled in F1-macro and Specificity (Missing and Tangling), suggesting strong performance across all classes. StableCode-3B's smallest storage requirements makes it a viable option for resource limited environments. GPT-4o performed best in the specificity sub-metrics Specificity (Exact) and Specificity (Tangling), indicating a strong capability in precisely aligning PRs. Finally, CodeGemma-7B proved most effective at identifying Missing PR-issue alignments.

The $R_{overall}$ metric identifies CodeLlama-7B as the best performing LLM for the fine-tuning task. Its $R_{overall}$ score of 1.42 and ranked first across fine-tuned LLMs showing a balanced and consistent proficiency across all measured performance categories, establishing it as the most effective model in this evaluation.


\endtcolorbox

%% file: results-interpretability-analysis.tex
\subsection{Interpretability Analysis Results}
\label{subsec:interpretability-analysis-results}

\begin{figure}[htbp]
    \centering
    \includegraphics[width=0.70\textwidth]{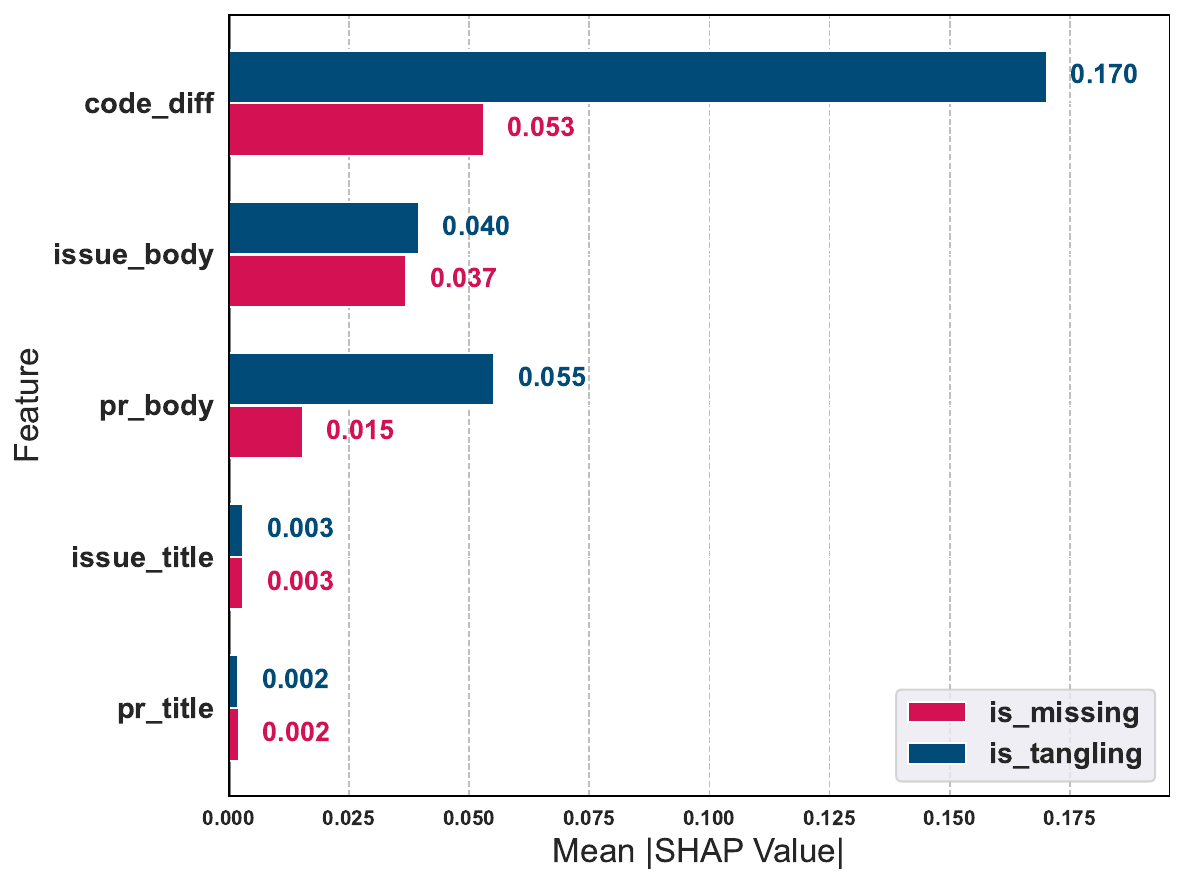}
    \caption{Global mean absolute SHAP values for \textit{isMissing} and \textit{isTangling} heads}
    \label{fig:shap_mean}
\end{figure}

\begin{figure}[htbp]
    \centering
    \includegraphics[width=0.75\textwidth]{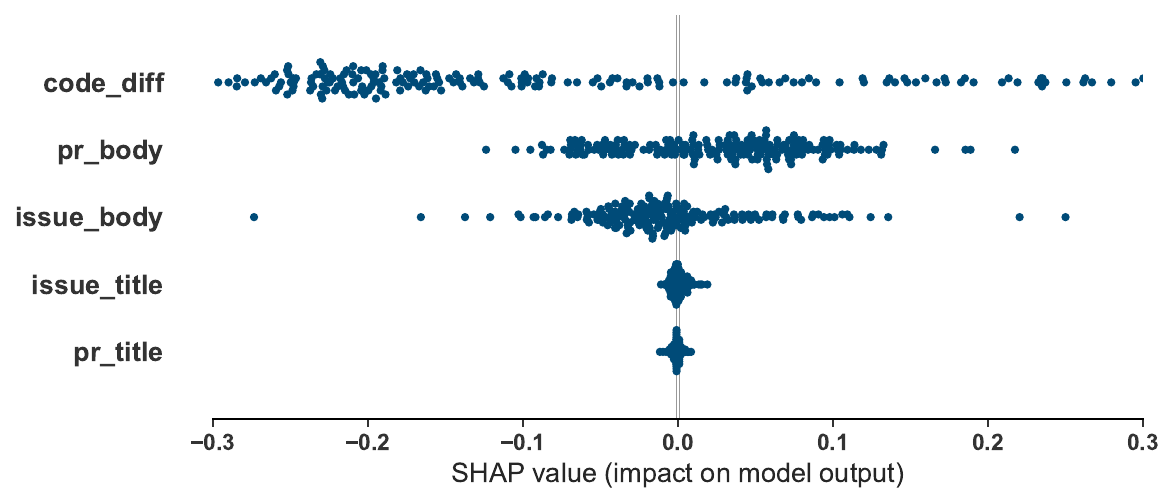}
    \caption{Beeswarm plot showing SHAP value distributions for the \textit{isTangling} head}
    \label{fig:shap_tangling_beeswarm}
\end{figure}

\begin{figure}[htbp]
    \centering
    \includegraphics[width=0.75\textwidth]{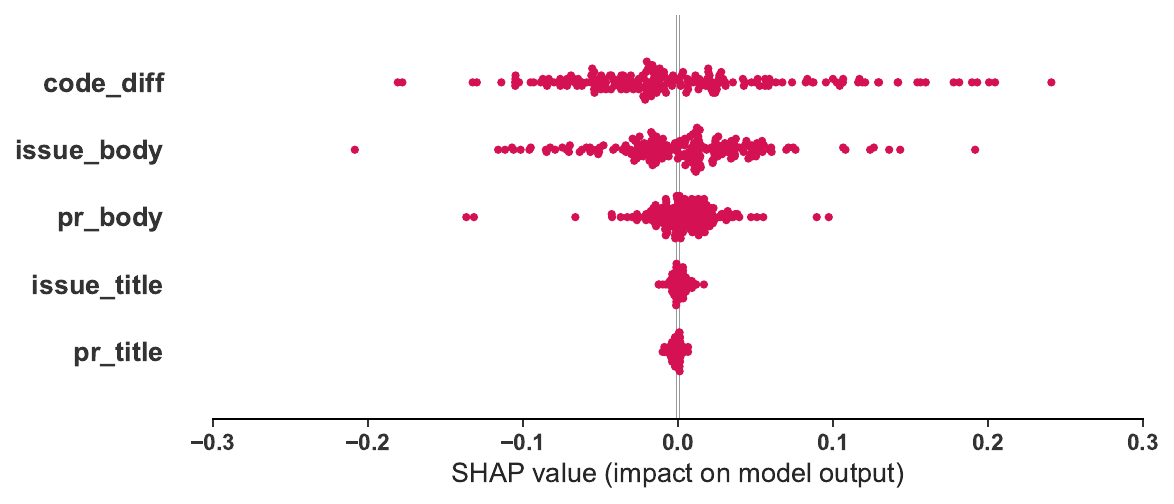}
    \caption{Beeswarm plot showing SHAP value distributions for the \textit{isMissing} head}
    \label{fig:shap_missing_beeswarm}
\end{figure}

We conducted SHAP-based interpretability analysis on the best-performing open-source LLM, CodeLLaMA-7B. We examined feature contributions separately for the \textit{isMissing} and \textit{isTangling} classification heads. Fig.~\ref{fig:shap_mean} presents the overall ranking of fields, while Fig.~\ref{fig:shap_missing_beeswarm} and  Fig. \ref{fig:shap_tangling_beeswarm} (beeswarm plots for missing and tangling heads) provide insight into how these contributions vary across individual predictions.  

\noindent\textbf{Global Feature Importances:} 
The global mean analysis (Fig.~\ref{fig:shap_mean}) highlights a consistent pattern across both heads: the code diff is the most influential field, though its dominance varies by task. The pr\_body and issue\_body provide secondary but meaningful signals, whereas titles pr\_title and issue\_title remain almost negligible.  

\noindent\textbf{\textit{isTangling} Head:} The model strongly prioritizes the \texttt{code\_diff}, which reaches a mean absolute SHAP value of 0.170. This value is nearly three times greater than pr\_body, the pull request body 0.055, and almost four times greater than the issue body 0.040. Titles exhibit near-zero contribution. (\texttt{issue\_title} $\approx$ \textbf{0.003}, \texttt{pr\_title} $\approx$ \textbf{0.002}). This ranking demonstrates that tangling is overwhelmingly a \textbf{code diff centric phenomenon}: the model primarily detects bundled functionality through structural patterns in the code changes.  
The relatively higher importance of the PR body compared to the issue body suggests that the model occasionally leverages descriptive text about the developer’s intent to reinforce code diff based evidence. For example, when the PR body discusses multiple tasks or touches on diverse concerns, this context can align with structural cues in the code diff, making tangling more evident. Nevertheless, the dominance of the code diff-with nearly triple the weight of the PR body---confirms that structural signals remain the decisive factor, while textual descriptions serve only as secondary support.  

\noindent\textbf{\textit{isMissing} Head:} In contrast, missing classification relies on a \textbf{joint signal}. The \texttt{code\_diff} contributes strongly (\textbf{0.053}), but the issue body is nearly as influential (\textbf{0.037}). This pairing reflects the nature of the task: the code diff provides the ``actual'' implementation, while the issue body provides the ``expected'' functionality. The PR body plays a smaller but still non-trivial role (\textbf{0.015}), while titles are again negligible (\texttt{issue\_title} $\approx$ \textbf{0.003}, \texttt{pr\_title} $\approx$ \textbf{0.002}). These values highlight that missing predictions are informed by both structural and textual fields together, rather than being dominated by a single source.  

\noindent\textit{Beeswarm Diagnostics.}  
The beeswarm plots provide a complementary, per-sample view of how features influence predictions.  

\noindent\textbf{Tangling beeswarm (Fig.~\ref{fig:shap_tangling_beeswarm})} The \texttt{code\_diff} exhibits by far the widest dispersion of SHAP contributions, with some samples showing extremely large positive or negative impacts. This confirms that tangling predictions are highly code diff-driven at the individual level as well. The PR body shows moderate variability, indicating that in certain cases the developer’s explanation provides additional context, but its spread is clearly smaller and far less consistent than that of the diff. The issue body contributes weakly, and titles remain tightly clustered near zero, offering virtually no predictive signal.  

\noindent\textbf{Missing beeswarm (Fig.~\ref{fig:shap_missing_beeswarm})} 
The distribution is more balanced: both \texttt{code\_diff} and \texttt{issue\_body} display moderate to wide spreads of SHAP values. This suggests that the model actively considers both what the issue describes and what the code implements when forming predictions. When the issue body lists requirements absent from the code diff, its SHAP contribution is strongly positive, pushing toward a missing classification. Conversely, if the code diff matches the issue description, its contribution may turn negative, signaling ``no missing.'' The PR body shows occasional moderate impact, while titles are clustered near zero.


\tcolorbox[colback=figure-soft-white,    
           colframe=figure-navy-blue] 

\textbf{RQ3: Which PR-issue fields exert the greatest influence on classification prediction?}

The interpretability analysis shows that the code diff is the single most influential field overall, but its role differs across classification heads. For \textit{isTangling} head, the importance ranking is: code diff (0.170) $\gg$ PR body (0.055) $>$ issue body (0.040) $>$  titles ($\approx$0.002--0.003). For \textit{isMissing} head, the ranking is more balanced: code diff(0.053) and issue body (0.037) jointly dominate, followed by PR body (0.015), with titles ($\approx$0.002--0.003) consistently negligible.  

These findings reveal three central insights. First, tangling is a code diff-centric phenomenon, with structural cues in the diff decisively shaping predictions, while textual bodies play only a supporting role. Second, missing detection relies on the joint signal of the code diff and the issue body, reflecting the need to consider both implementation and requirements together. Third, titles add negligible value across both tasks, suggesting that their limited semantic content makes them unhelpful for classification in this setting.

\endtcolorbox

%% file: threats-to-validity.tex
\section{Threats to Validity}
\label{sec:threats-to-validity}

This Section outlines potential threats to internal, construct, and external validity. Internal threats include labeling errors, decision bias and the assumption of issue atomicity. Construct threats arise from missing implicit requirements and external resources. External validity is limited by dataset size and scope, affecting the generalizability across repositories and languages.

\subsection{Threats to Internal Validity}
\label{subsec:threats-to-internal-validity}

A potential threat to the internal validity of our findings stems from the manual labeling of PR-issue pairs. The labeling process, which classified PRs based on their association with specific issues, was conducted by independently by two authors, with a third author arbitrating any disagreements. Although this approach was designed to foster objectivity and a detailed replication package was created to document the rationale for each label, the inherent complexity of code repositories introduces a risk of human error and subjective interpretation. Raters, despite their expertise, may not always have a complete view of the system's architecture, which can make it difficult to fully evaluate the broader implications of a code change. While this creates a potential source of oversight or bias, our structured evaluation protocol and arbitration process were specifically designed to minimize these risks.

\begin{figure}[t]
\centering
\includegraphics[width=\columnwidth]{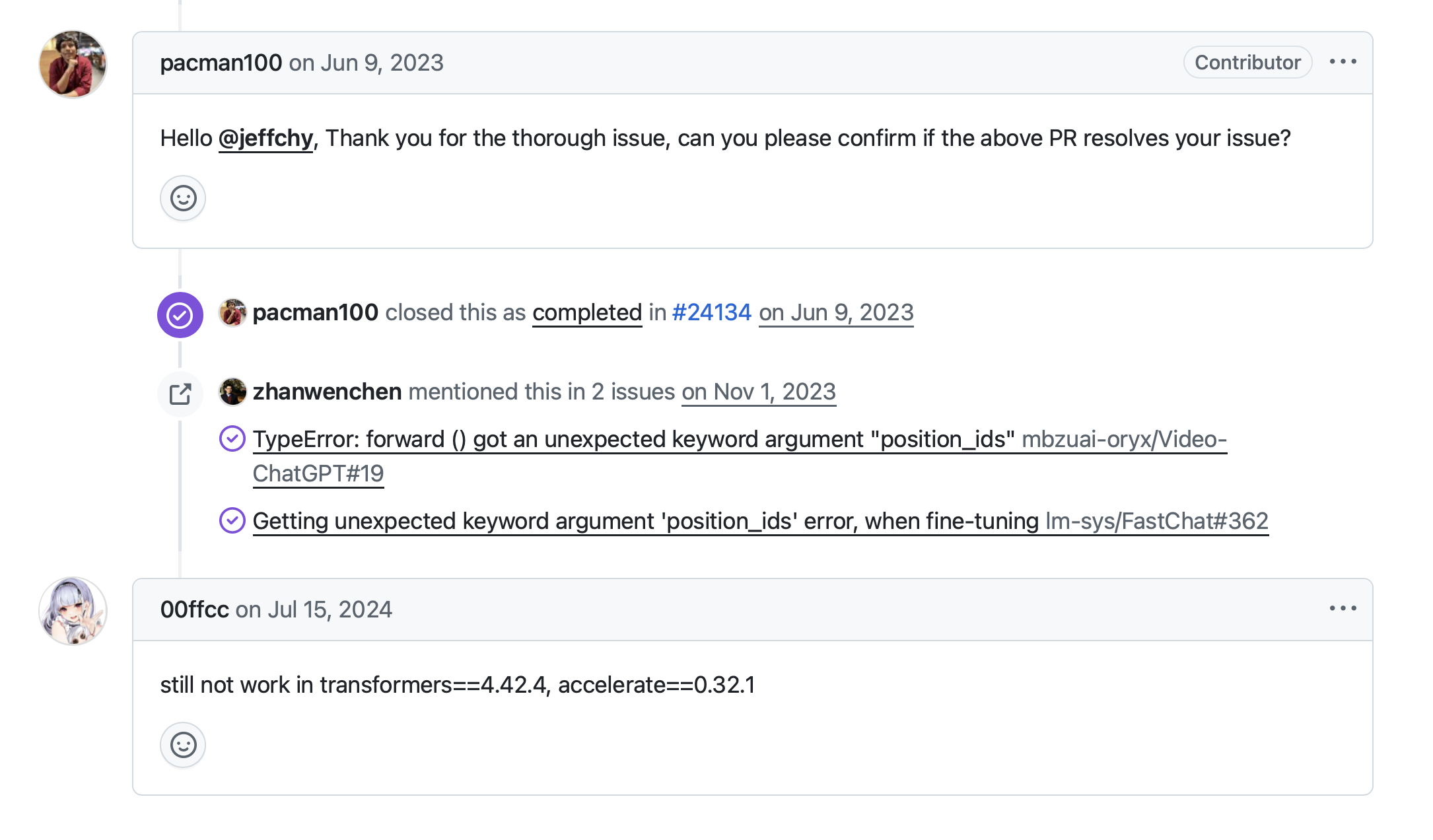}
\caption{Issue discussion inducing confirmation bias}
\label{fig:internal_threat_2}
\end{figure}

A second threat to internal validity concerns potential decision bias during PR evaluation. Comments within PR and issue threads may inadvertently influence annotators toward specific classifications, even when the evidence is ambiguous. For instance, Fig. \ref{fig:internal_threat_2} illustrates a pull request initially merged as a resolution, later questioned by another developer. These contextual cues can introduce confirmation bias. Our evaluation protocol, involving a two-stage assessment with independent judgments and subsequent reconciliation by three raters, mitigates the influence of any single perspective. While this approach substantially reduces the impact of such bias, it may still persist in rare cases where contextual information is particularly strong or ambiguous.

\begin{figure}[t]
\centering
\includegraphics[width=\columnwidth]{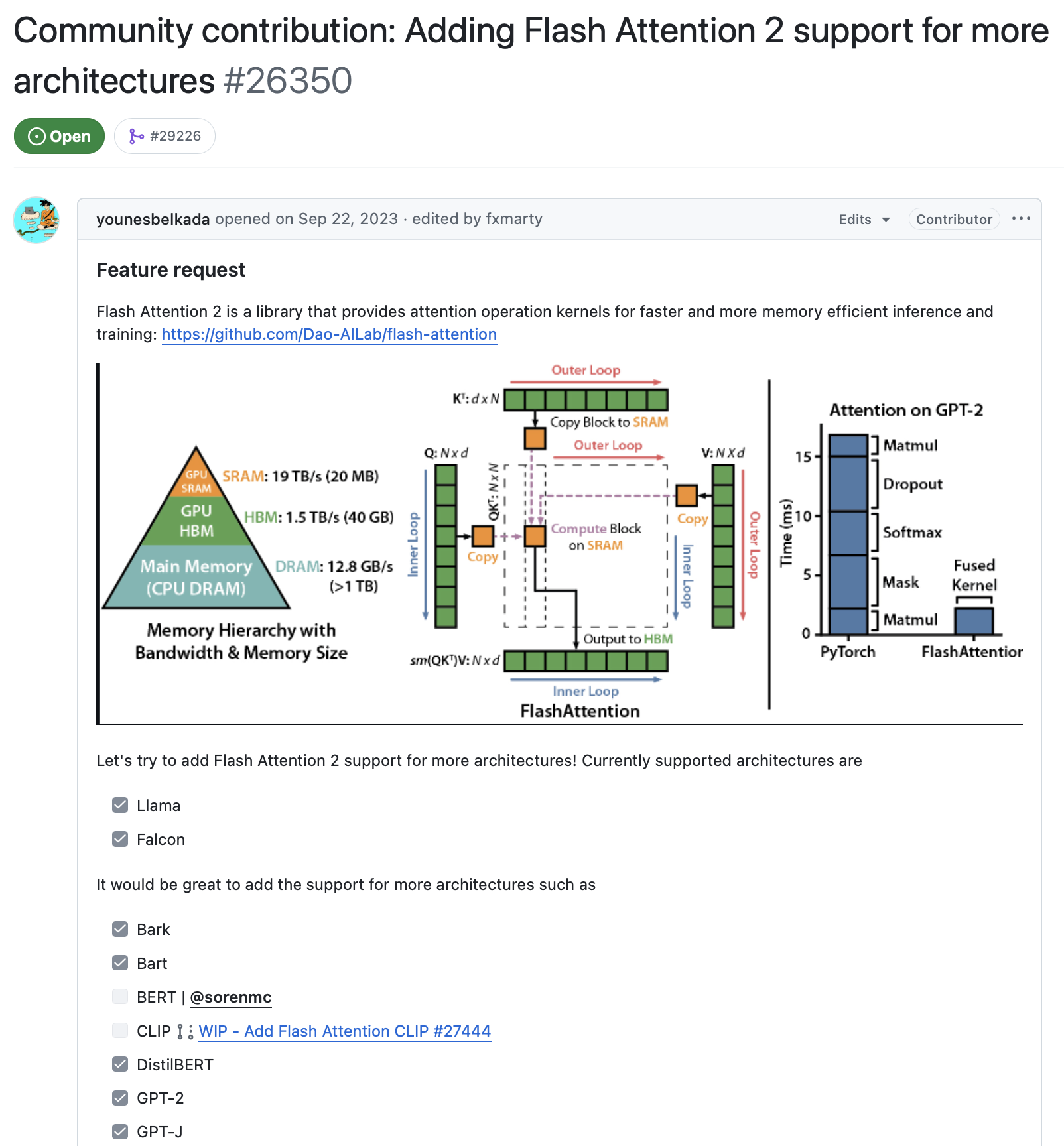}
\caption{An example of a non-atomic issue.}
\label{fig:internal_threat_4}
\end{figure}

A third threat to internal validity is based on our foundational assumption for the PR-issue alignment problem. Recent taxonomy assumes that if an issue is atomic ($A$), there exists a pull request that can be assigned one of four alignment categories: $Exact$, $Missing$, $Tangling$, or $Missing \& Tangling$. While it is generally considered best practice for issues to be atomic, there may exist rare cases in software development where issues are not strictly atomic. In such edge cases, the alignment taxonomy may provide only an approximate reflection of the true nature of the work. As shown in Fig.~\ref{fig:internal_threat_4}, an issue requests a single feature to be implemented across multiple models, which represents one of these exceptional situations.

\subsection{Threats to Construct Validity}
\label{subsec:threats-to-construct-validity}

\begin{figure}[htbp]
    \centering
    \begin{subfigure}[t]{0.48\textwidth}
        \centering
        \includegraphics[width=\linewidth]{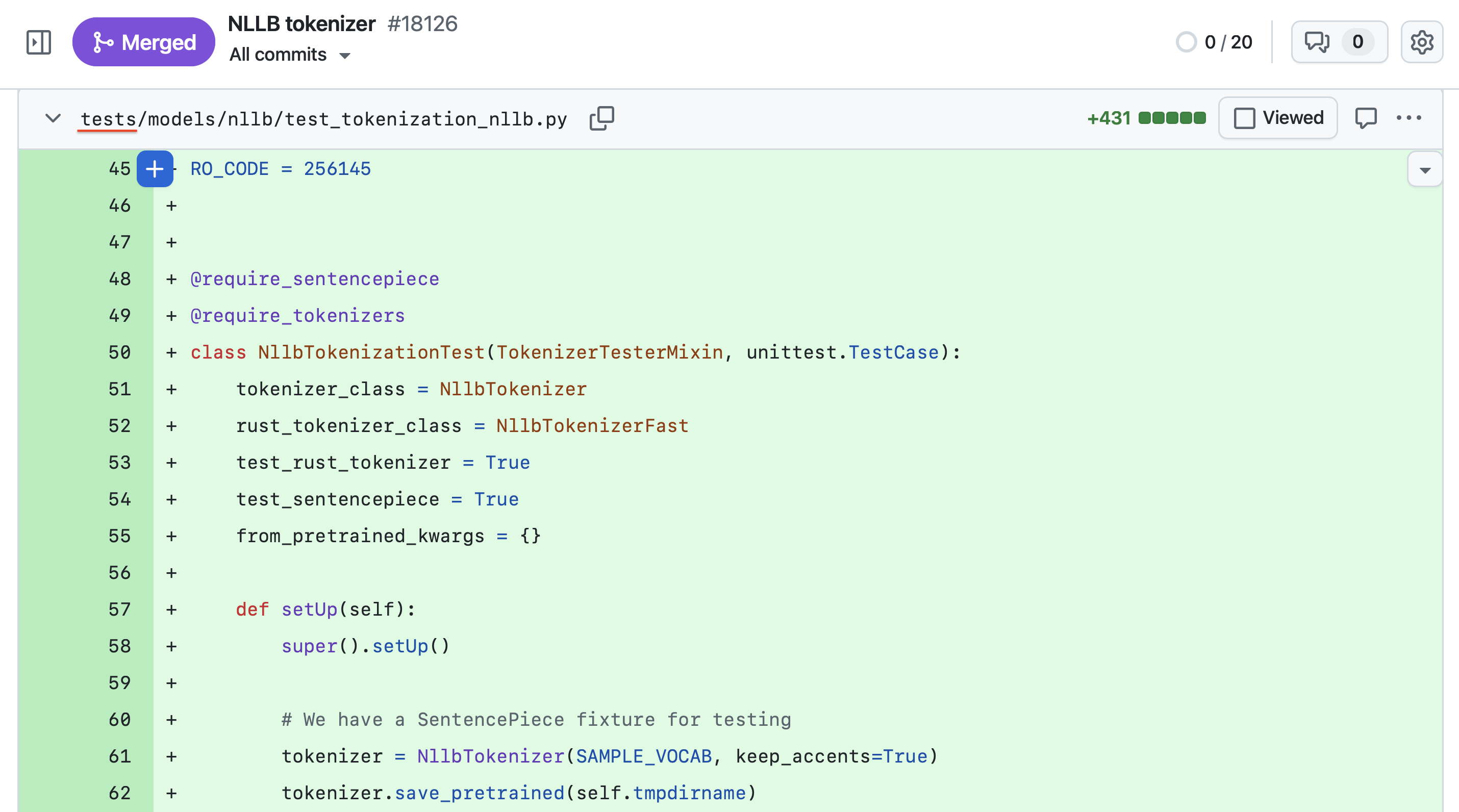}
        \caption{A PR with tests not explicitly requested}
        \label{fig:construct_threat_1_1}
    \end{subfigure}
    \hfill
    \begin{subfigure}[t]{0.48\textwidth}
        \centering
        \includegraphics[width=\linewidth]{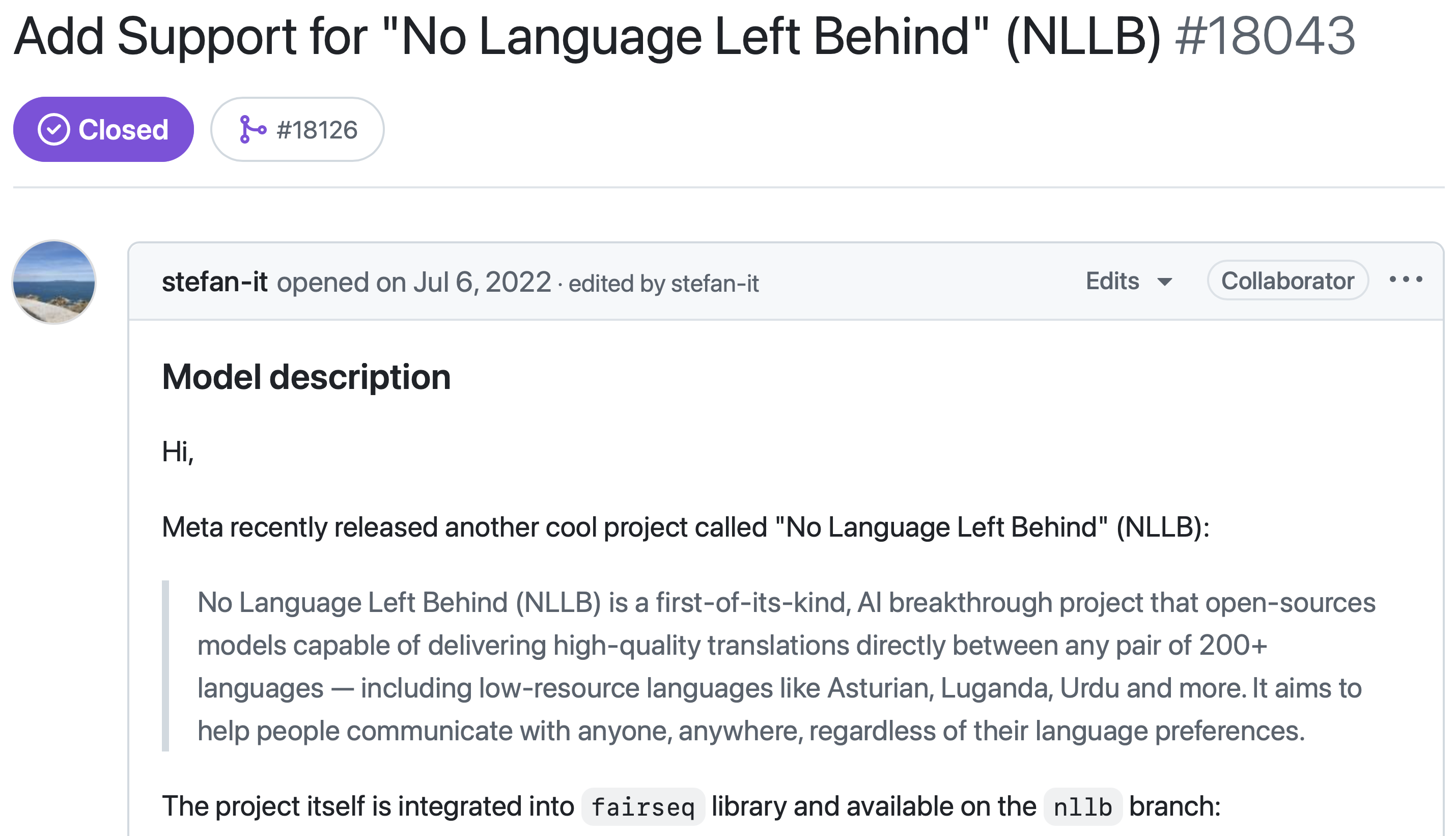}
        \caption{The issue not asking for tests}
        \label{fig:construct_threat_1_2}
    \end{subfigure}

    \vspace{0.5em} 
    \begin{subfigure}[t]{0.95\textwidth}
        \centering
        \includegraphics[width=\linewidth]{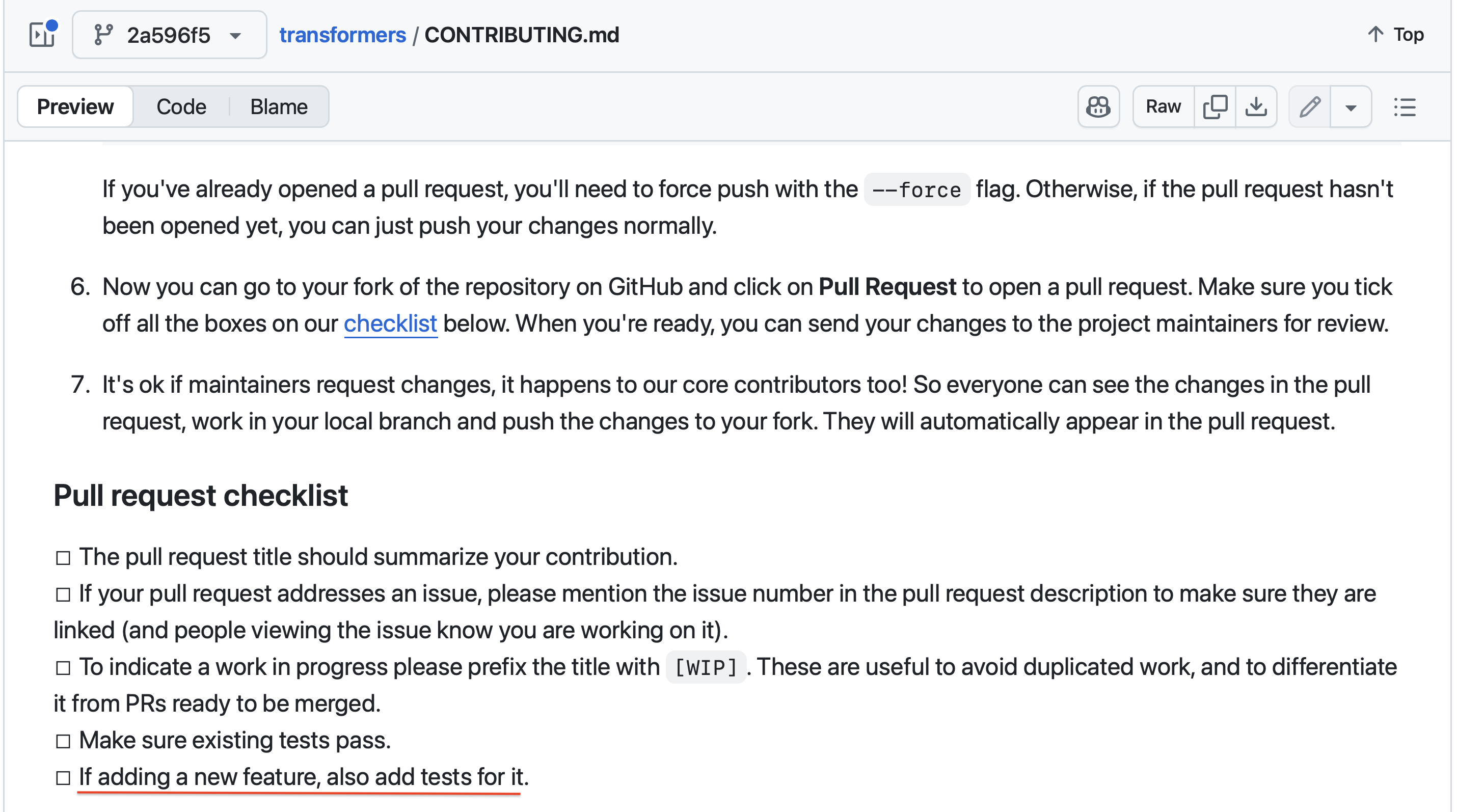}
        \caption{The project's guidelines requiring tests}
        \label{fig:construct_threat_1_3}
    \end{subfigure}

    \caption{The project's guidelines requiring tests}
    \label{fig:construct_threat_1}
\end{figure}

A potential threat to the construct validity of our study arises from the failure to capture all relevant dimensions of the PR-issue alignment problem. The PR titles and issue descriptions, which serve as the primary inputs for our model, may not explicitly represent all implicit requirements of a software project. A key example of this is a project's community contribution guidelines. For instance, while it is implicitly expected that all code contributions include relevant test methods, this requirement is often not explicitly stated in the issue or pull request fields. This is demonstrated in Fig.~\ref{fig:construct_threat_1}, where a PR shown in Fig.~\ref{fig:construct_threat_1_1} adds test methods even though the corresponding issue in Fig.~\ref{fig:construct_threat_1_2} does not explicitly request them. This behavior is consistent with the project's contribution guidelines shown in Fig.~\ref{fig:construct_threat_1_3}, which require new code to have corresponding tests. Since our model is not provided with these external guidelines, it lacks the necessary information to fully capture the true alignment of the PR-issue pair.

\begin{figure}[htbp]
\centering
\includegraphics[width=\columnwidth]{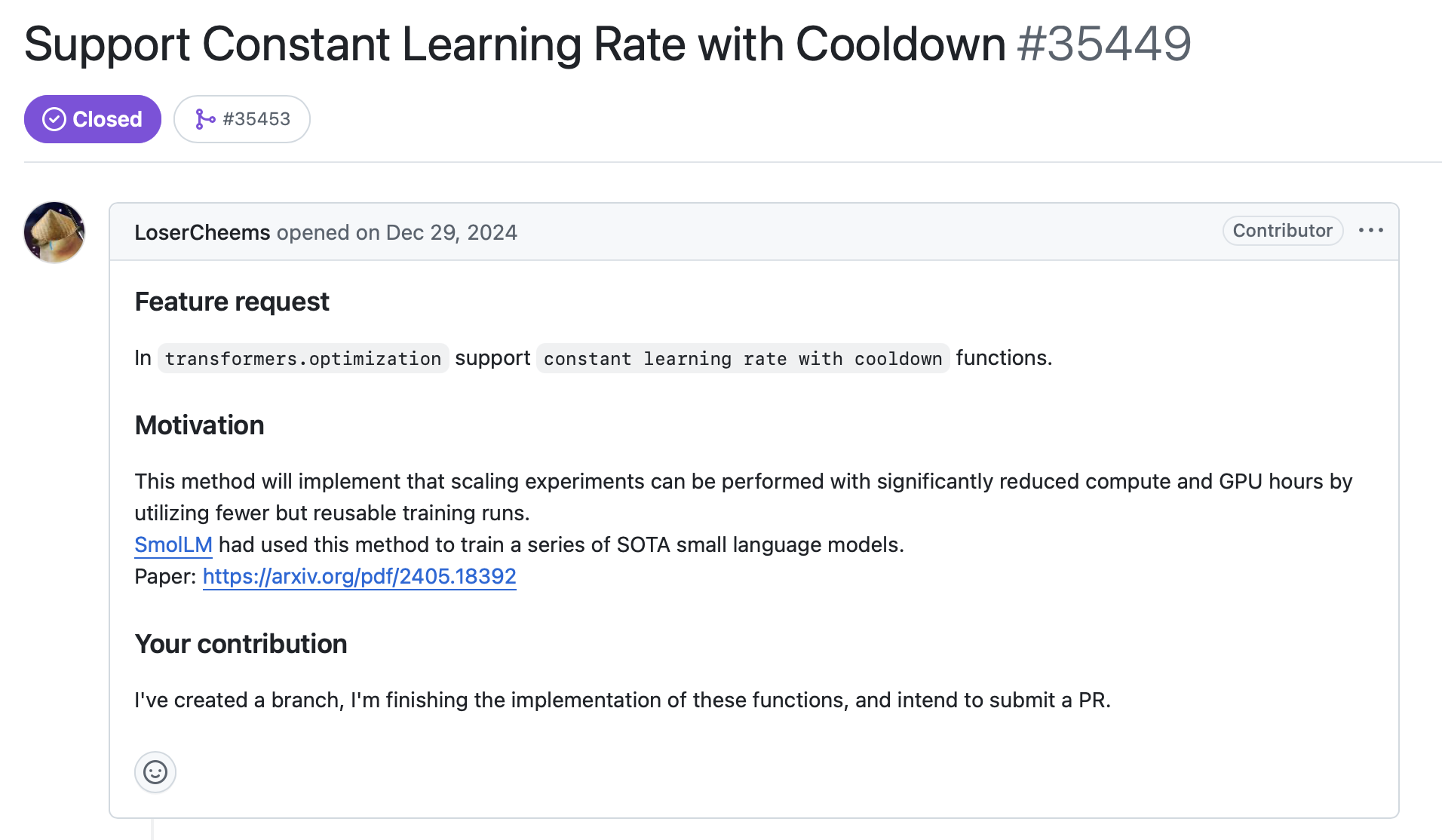}
\caption{Issue with external resources}
\label{fig:construct_threat_2}
\end{figure}

A second threat to construct validity arises from the not capturing of external resources from our analysis. Some issues and PRs are not self-contained and require external context, such as links to research papers, design documents, or other websites, to be fully understood. As shown in Fig.~\ref{fig:construct_threat_2}, an issue may ask a developer to implement a feature based on a specific external resource. Because our LLM is not equipped to process or access these external sources, it is unable to accurately evaluate whether a given PR correctly addresses the requirements specified within the linked document. This limitation means our model may misclassify a PR.

\subsection{Threats to External Validity}
\label{subsec:threats-to-extenal-validity}
A potential threat to the external validity of our study arises from the characteristics of our dataset. Due to the labor-intensive nature of manual labeling, our analysis was conducted on a limited, stratified sample from the selected repository. While this approach preserved the proportions of merged and unmerged pull requests, the small sample size and its exclusive origin from a single repository pose a significant limitation. Furthermore, selected repository primarily uses Python, which means our findings may not be generalizable to projects with different programming languages or those from other domains. These constraints on the dataset's scope may limit the broader applicability of our conclusions to the wider landscape of software development.

%% file: discussion.tex
\section{Discussion}
\label{sec:discussion}

This Section highlights the practical usability of our findings and directions for future research. Practically, the results support tool development for PR–issue alignment and integration into workflows. Research extensions include testing alternative fine-tuning methods, enriching context, refining label mapping, and assessing performance in cross-project and multilingual settings. 

\subsection{Implication for Practitioners}

\textbf{Future Tool Development:} The development of software tools leveraging fine-tuned LLMs for PR-issue alignment classification is now more feasible due to their significantly reduced storage requirements and improved performance compared to previous methods. Consequently, these models can be deployed and executed locally on edge devices, without the need for large server infrastructure, enabling more flexible and decentralized usage scenarios. Such tools could be employed to systematically evaluate developers PR characteristics, inform decision-making processes, and provide automated support in project management. Additionally, by complementing existing code review bots, Agentic AI that classifies PR–issues can provide additional context, improving workflow efficiency and maintaining traceability between PRs and issues within integrated development environments. \\

\noindent\textbf{Workflow Integration:} Recent PR-issue (Exact, Missing, Tangling, Missing and Tangling) alignment taxonomy and fine-tuned LLM models can be effectively integrated into existing software development pipelines, allowing automated prioritization of PR reviews based on alignment quality. This integration can enhance traceability and consistency across development tasks, streamline code review processes, and provide actionable insights to development teams, ultimately supporting more structured and efficient software maintenance workflows. \\

\noindent\textbf{Code Review Tool Integration:}  The four-class PR–issue alignment taxonomy (Exact, Missing, Tangling, Missing and Tangling) and associated models can be integrated into existing code review tools, thereby transforming code review workflows in practice. While numerous code review bots exist, none currently incorporate PR–issue alignment classification, making this a complementary feature that enriches these tools with additional context. By embedding these models into the code review process, practitioners can receive automated and systematic insights on potential misalignment or tangling issues early on, reducing manual cognitive effort and enhancing review quality. Such integration has the potential to streamline feedback cycles, improve decision-making, and ultimately foster more efficient and consistent code reviews across teams. Moreover, aligning the proposed taxonomy with current development pipelines could help practitioners adopt more evidence-driven practices, ensuring that the review process is not only more reliable but also more scalable to complex, large-scale software projects.\\

\noindent\textbf{Reflection of Coding Practices:} The developed models and taxonomy can be systematically integrated into repository-level metrics to provide practitioners with structured feedback on their software engineering practices. By moving beyond traditional productivity indicators, such integration can surface patterns of PR–issue misalignment and code tangling that may otherwise remain hidden. This reflective layer of analysis would not only raise awareness among developers and teams about their coding and collaboration practices but also encourage critical evaluation of existing workflows. In turn, such feedback mechanisms could foster discussions on improving software quality, aligning contributions more effectively with project goals, and questioning entrenched practices that may undermine long-term maintainability.

\subsection{Implication for Researchers}

\textbf{Testing other Fine-Tuning Methods:} Future research can explore and evaluate the effects of alternative fine-tuning strategies, including full fine-tuning, LoRA, QLoRA, and Adapter-Tuning, on PR-issue alignment classification performance. Systematic investigation of these methods can provide insights into how different fine-tuning processes influence model accuracy, efficiency, and generalizability, offering guidance for selecting optimal strategies for domain-specific classification tasks. Such studies can also inform best practices for balancing resource requirements with classification effectiveness in fine-tuning LLMs.\\

\noindent\textbf{Exploring Alternative Data Augmentation Strategies:} Future research can also investigate the impact of various data augmentation techniques on PR-issue alignment classification performance. Methods such as back-translation, synonym replacement, random insertion, deletion, and LLM text augmentation with few-shot prompting  can be systematically evaluated to determine their effects on improving model generalization and accuracy. By examining how these augmentation strategies influence fine-tuned LLMs across different alignment categories, researchers can identify effective ways to mitigate data imbalance, enhance representational diversity, and improve classification accuracy in domain-specific software engineering contexts.\\
    
\noindent\textbf{Context Enrichment:} Future research can enhance LLM-based PR-issue alignment models by incorporating richer contextual information from PR and issue threads. Extending model inputs to include discussions, comments, and linked documentation can improve understanding of implicit requirements that are not explicitly captured in titles or descriptions. Moreover, agentic and multimodal systems can be employed to retrieve, summarize, and interpret external documents, as well as explain visual artifacts, thereby providing a more comprehensive context for classification. Such approaches can address construct validity concerns by ensuring that models consider the full range of project-specific expectations and guidelines, ultimately improving alignment accuracy and decision-making support. \\

\noindent\textbf{Confidence Bounds in Label Mapping:}  In certain edge cases, determining the exact PR-issue alignment class can be challenging, highlighting the need for more nuanced labeling strategies. Future work can enhance PR-issue alignment datasets by incorporating confidence levels for labels, with relaxed boundaries (e.g., between 0.4 and 0.6 considered uncertain) and clearly defined confident ranges. This probabilistic approach allows annotators to better handle borderline cases and provides more reliable guidance for automated large-scale evaluations, ultimately improving dataset robustness and the credibility of model performance metrics. \\
    
\noindent\textbf{Determination of Optimal Label Thresholds:} 
Building on the previous implication, rather than arbitrarily selecting label mapping thresholds (e.g., 0.5), future research can employ \textit{data-driven calibration techniques} to determine optimal cutoffs for probabilistic label assignments. By systematically analyzing model confidence distributions and evaluating trade-offs across precision–recall or ROC curves, these methods can identify threshold values that maximize classification performance and reduce uncertainty in borderline cases. Such threshold optimization provides a principled framework for improving reliability and standardization in PR–issue alignment classification. \\

\noindent\textbf{Cross-Project Generalizability:} Further research is required to evaluate the generalizability of PR-issue alignment models across diverse programming languages, repositories, and development contexts. Investigating how different domain adaptation strategies, such as translating code diffs or project-specific conventions, influence model performance can provide insights into cross-project applicability. Such studies can inform best practices for transferring fine-tuned LLMs between projects, ultimately supporting more robust and versatile alignment models in heterogeneous software engineering environments. \\



%% file: conclusion.tex
\section{Conclusion}
\label{sec:conclusion}

PR-based development remains hindered by misalignments between PRs and their associated issues, undermining traceability, maintainability, and defect localization. Prior research has focused on detecting and untangling tangling commits but has underutilized fine-tuned LLMs for comprehensive classification across a more nuanced PR–issue alignment taxonomy. This study fills that gap by fine-tuning open-source LLMs, showing they outperform prompt-based methods in capturing domain-specific distinctions. The contributions are: (1) fine-tuning closed-source and open-source LLMs for PR–issue classification, (2) identifying the best performing fine-tuned LLM for PR-issue classification (3) revealing the most influential PR–issue fields, and (4) proposing a novel two classification head LLM-based classification architecture.

Fine-tuned LLMs outperform baseline and proprietary models, offering more reliable PR–issue alignment classification. CodeLlama-7B proved most consistent overall, while Deepseek-Coder-6.7B, ChatGPT-4o, and CodeGemma-7B each excelled in specialized tasks. StableCode-3B showed promise for resource-limited contexts. Interpretability analysis highlighted code diffs and issue  descriptions as key factors, emphasizing the combined role of technical and textual contexts. These results confirm the effectiveness of fine-tuning and clarify how models and input fields shape classification quality.

Looking forward, our study highlights key implications and limitations. Fine-tuned models, with lower storage demands and improved performance, facilitate workflow integration for automated alignment checks and code review prioritization. Beyond workflow integration, the developed taxonomy and models can also be embedded into code review tools and repository-level metrics, offering practitioners structured feedback that enhances code review quality and raises awareness of coding practices. Researchers can explore alternative fine-tuning strategies, PeFT, and model trade-offs, while extending contextual inputs to discussion threads, linked commits, and documentation. Limitations include issue atomicity assumptions, and a dataset limited to a single repository, which constrains generalizability. Refining label mappings and evaluation frameworks could further strengthen methodological rigor, supporting future research and experimentation in LLM adaptation for software engineering.

This study demonstrates how fine-tuned LLMs can substantially improve PR–issue alignment classification and highlights which PR–issue fields most influence model decisions. By fine-tuning  LLMs, our work establishes a robust foundation for researchers and practitioners to advance PR-issue alignment classification, automated code review, contextual understanding. We hope this work inspires both practitioners and researchers to rethink  how they approach to software analysis, moving beyond traditional methods to address the deeper, systemic challenges shaping contemporary software development workflows in the era of AI.\\

